\documentclass[12pt]{article} 

\newcommand{\blind}{0}

\usepackage{times}
\usepackage{graphicx,psfrag,epsf}
\usepackage{amsmath}
\usepackage{amssymb}
\usepackage{gensymb}
\usepackage{subcaption}
\usepackage[toc,page]{appendix}
\usepackage{natbib}
\usepackage{xcolor}

\usepackage{graphicx} 

\usepackage{url}
\usepackage{booktabs}
\usepackage{array}
\usepackage{multirow}

\newtheorem{theorem}{Proposition}

\usepackage{enumitem}

\usepackage{setspace}

\usepackage{verbatim}
\usepackage[format=plain, labelfont=bf, singlelinecheck=off]{caption} 

\usepackage{algorithm}
\usepackage{algpseudocode}
\newlength{\continueindent}
\usepackage{etoolbox}
\makeatletter
\newcommand*{\ALG@customparshape}{\parshape 2 \leftmargin \linewidth \dimexpr\ALG@tlm+\continueindent\relax \dimexpr\linewidth+\leftmargin-\ALG@tlm-\continueindent\relax}
\apptocmd{\ALG@beginblock}{\ALG@customparshape}{}{\errmessage{failed to patch}}
\makeatother

\numberwithin{equation}{section}

\def\bc{\textbf {c}}

\def\bu{\textbf {v}}

\def\bD{\textbf {D}}
\def\bA{\textbf {A}}
\def\bB{\textbf {B}}

\def\bK{\textbf {K}}
\def\bI{\textbf {I}}

\def\bQ{\textbf {Q}}
\def\bs{\textbf {s}}
\def\bH{\textbf {H}}

\def\bu{\textbf {u}}

\def\bS{\textbf {S}}
\def\bV{\textbf {V}}

\def\bX{\textbf {X}}
\def\bY{\textbf {Y}}

\def\bepsilon{\boldsymbol {\epsilon}}
\def\boeta{\boldsymbol {\eta}}
\def\bdelta{\boldsymbol {\delta}}
\def\bSigma{\boldsymbol {\Sigma}}

\def\bmu{\boldsymbol {\mu}}

\def\bbeta{\boldsymbol {\beta}}

\def\bxi{\boldsymbol {\xi}}
\def\bfxi{\boldsymbol {\xi}}
\def\btheta{\boldsymbol {\theta}}

\def\bfzero{\mbox{\boldmath $0$}}

\def\bfSigma{\mbox{\boldmath $\Sigma$}}
\def\var{\mbox{\textrm{var}}}

\def\cov{\mbox{\textrm{cov}}}

\expandafter\def\expandafter\normalsize\expandafter{%
    \normalsize
    \setlength\abovedisplayskip{6pt}
    \setlength\belowdisplayskip{6pt}
    \setlength\abovedisplayshortskip{6pt}
    \setlength\belowdisplayshortskip{6pt}
}

\usepackage{mathtools}

\DeclarePairedDelimiter\floor{\lfloor}{\rfloor}

\begin{document}
\date{}

\def\spacingset#1{\renewcommand{\baselinestretch}%
{#1}\small\normalsize} \spacingset{1}


\if0\blind
{
  \title{\bf Spatial Statistical Downscaling for Constructing High-Resolution Nature Runs in Global Observing System Simulation Experiments 
  }

  \author{Pulong Ma \\
  The Statistical and Applied Mathematical Sciences Institute  \\
  and Duke University \\ 
    Emily L. Kang \\
    Department of Mathematical Sciences, University of Cincinnati\\
    Amy J. Braverman 
    and
    Hai M. Nguyen\\
     Jet Propulsion Laboratory, California Institute of Technology
    }    

  \maketitle

} \fi

\if1\blind
{
  \bigskip
  \bigskip
  \bigskip
  \begin{center}
    {\LARGE\bf Spatial Statistical Downscaling for Constructing High-Resolution Nature Runs in Global Observing System Simulation Experiments}
\end{center}
  \medskip
} \fi

\bigskip
\begin{abstract}
Observing system simulation experiments (OSSEs) have been widely used as a rigorous and cost-effective way to guide development of new observing systems, and to evaluate the performance of new data assimilation algorithms. Nature runs (NRs), which are outputs from deterministic models, play an essential role in building OSSE systems for global atmospheric processes because they are used both to create synthetic observations at high spatial resolution, and to represent the ``true'' atmosphere against which the forecasts are verified. However, most NRs are generated at resolutions coarser than actual observations from satellite instruments or  predictions from data assimilation algorithms. Our goal is to develop a principled statistical downscaling framework to construct high-resolution NRs via conditional simulation from coarse-resolution numerical model output. We use nonstationary spatial covariance function models that have basis function representations to capture spatial variability. This approach not only explicitly addresses the change-of-support problem, but also allows fast computation with large volumes of numerical model output. We also propose a data-driven algorithm to select the required basis functions adaptively, in order to increase the flexibility of our nonstationary covariance function models. In this article we demonstrate these techniques by downscaling a coarse-resolution physical numerical model output at a native resolution of $1^{\circ} \text{ latitude} \times 1.25^{\circ} \text{ longitude}$ of  global surface $\text{CO}_2$ concentrations to 655,362 equal-area hexagons.
\end{abstract}

\noindent%
\vfill

\newpage
\spacingset{1.45} 

\section{Introduction} \label{sec: intro}

\begin{figure}[htbp]
\begin{center}
\makebox[\textwidth][c]{ \includegraphics[width=.95\textwidth, height=0.2\textheight]{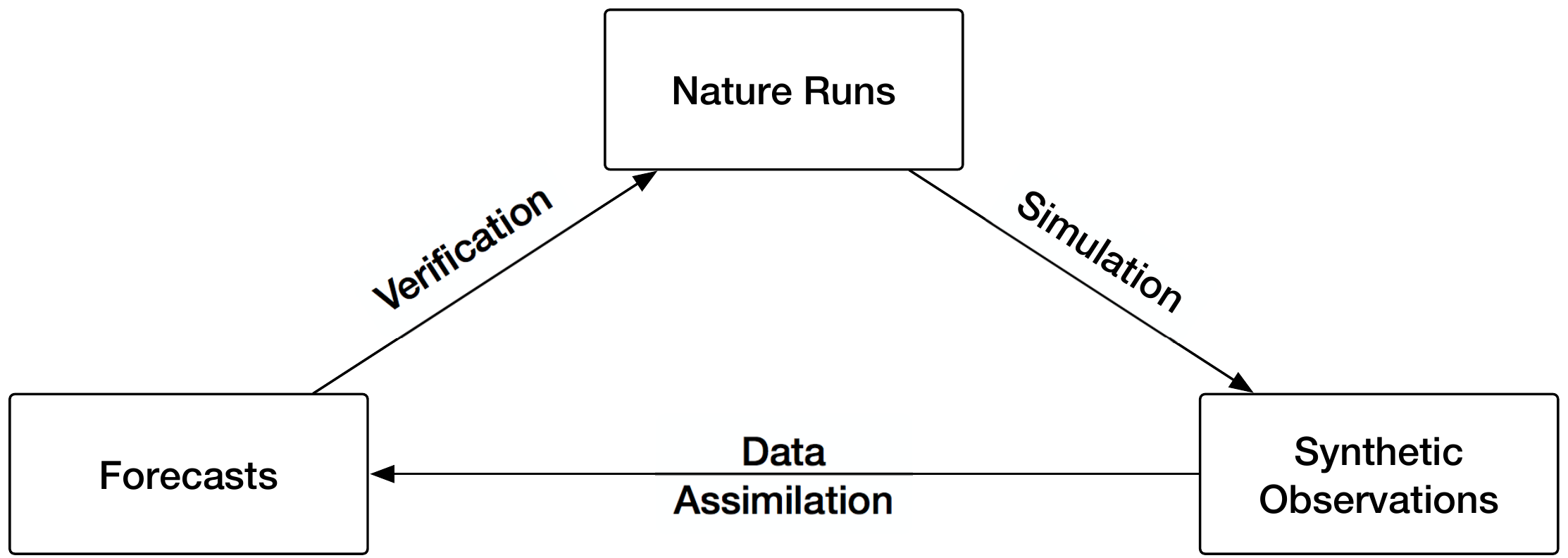}}
\caption{An OSSE system. The ``nature runs'' (NRs) are generated by or downscaled from numerical model outputs, and represent the assumed ``true" state of the atmosphere.  Synthetic observations are simulated by adding error components to NRs to mimic realistic cloud coverage and measurement errors. Forecasts from data assimilation algorithms are compared with NRs to evaluate the predictive performance of these algorithms.  \vspace{-0.6cm}} 
\label{fig: osse}
\end{center}
\end{figure}

Observing system simulation experiments (OSSEs) are widely used in atmospheric studies and climate monitoring to guide development of new observing systems including satellite missions and ground-based monitoring networks, and to evaluate performance of new data assimilation algorithms \cite[e.g.,][]{Edwards2009, Zoogman2011, Errico2013, Atlas2015, Hoffman2016}; see Figure~\ref{fig: osse} for a diagram of a basic OSSE. In an OSSE, a simulated atmospheric field from a numerical model is used as the ``truth'' (termed a Nature Run, or NR) to produce synthetic observations by adding suitable measurement errors and other representative errors such as cloud mask \citep[e.,g.,][]{Atlas2015, Hoffman2016}. These synthetic observations are then fed to a data assimilation algorithm. Here, data assimilation refers to the process of fusing ground-based or airborne observations from observing systems such as satellites together with numerical model output, to infer the true state of geophysical processes; see \cite{Wikle2007} for a formal definition of data assimilation from a statistical perspective. The estimated true states are typically called forecasts in atmospheric sciences. Since OSSEs deal entirely with simulations, they provide a cost-effective approach to evaluating the impact of new observing systems and performance of new data assimilation algorithms, and can be used when actual observational data are not available. In particular, OSSEs can be employed to compare competing observing system designs \citep[e.g.,][]{Atlas2015, Hoffman2016}. Moreover, unlike comparison against in-situ observations, the ``truth'' in an OSSE (that is, the NR) is known and uncontaminated, and thus can be directly used to better determine the accuracy and precision of forecasts. OSSEs are also extremely useful in understanding and quantifying capabilities of new satellite mission designs. For instance, \cite{Abida2017} use OSSEs to evaluate the potential improvement in estimating global surface carbon monoxide with the proposed Geostationary Coastal and Air Pollution Events Mission (GEO-CAPE). \cite{Liu2017} use OSSEs to investigate the potential for high spatial resolution satellite NO$_2$ observations to estimate surface NO$_2$ emissions. 

The nature run (NR) is an essential component of an OSSE, since it provides the ``true'' state of the atmosphere, from which synthetic observations are constructed. As such, it provides a standard for evaluating the quality of forecasts from data assimilation algorithms (Figure~\ref{fig: osse}). For OSSEs to be useful, it is essential that their NRs reflect characteristics of the real atmosphere. Hence, NRs are often generated by state-of-the-art numerical atmospheric models driven by a set of ordinary and partial differential equations describing atmospheric chemistry and dynamics. When such a numerical model is run for the entire globe, or for a large geographical region, model complexity and computational limitations demand that many complex geophysical processes be simplified, which is referred to as parameterization \citep[e.g.,][pp.~342-398]{Brasseur2017}, and that the output be produced at low spatial and temporal resolution. On the other hand, with advancements in sensing technologies, new ground-based and space-based instruments are expected to provide observations with higher and higher spatial resolutions. Studies of local or regional air quality, or anthropogenic emissions and their impacts, require high spatial resolutions in order to be relevant for natural resource management and environmental policy decisions. As a result, NRs in OSSEs need to be generated at increasingly high spatial resolutions. The difference in resolution between numerical model output and that required to evaluate new sensor designs, or for local/regional scales analyses, motivates the need to construct finer resolution outputs from coarse-resolution numerical model outputs when generating NRs. This is called downscaling in remote sensing and atmospheric science \citep[e.g.,][]{Gutmann2012,Atkinson2013,Glotter2014}, and is an example of what is known as the  change-of-support problem in spatial statistics \citep{Cressie1993, Gotway2002, Banerjee2014}.

In the OSSE literature, heuristic methods are often used to downscale (sometimes called ``preprocess" ) model output to higher resolutions \citep[e.g.,][]{Eskes2003, Errico2013,Tsai2014}. These methods are computationally fast, but generally \emph{ad hoc}. The NASA Global Modeling and Assimilation Office (GMAO) has run a global non-hydrostatic dynamical core to perform cloud-system resolving experiments at resolutions as fine as 3.5 km within the NASA Goddard Earth Observing System global atmospheric model version 5 (GEOS-5) \citep{Putman2011}. Simulations from this study provide high resolution NRs for geophysical variables, but they are computationally expensive. For example, generating two-year GEOS-5 simulations requires 61 days of computing with 7,200 cores and a total of $10.5\times 10^6$ core hours.  These requirements make it difficult for such an approach to be used widely \citep{Webster2015}.

There is a vast literature on statistical downscaling and its applications in atmospheric studies. Most of this is devoted to developing methods for comparing, correcting, or calibrating numerical model output  using observed data from physical experiments and monitoring stations \cite[e.g.,][]{Lenderink2007, Kloog2011, Zhou2011, Berrocal2012, Reich2014}. These methods require \textit{both} numerical model output and adequate in-situ observations \textit{together} in order to fit a statistical model that relates coarse-resolution model outputs to observations.  Without explicitly dealing with the change-of-support problem, many methods use observations as response variables in simple linear regressions (or some variants), with the model outputs as explanatory variables. For instance, \cite{Guilla2008} develop a two-step linear regression procedure to downscale numerical model outputs, and adjust them to fit monitoring station data. There are also methods to extend these downscaling ideas to handle multiple variables in space and time \citep[e.g.,][]{Berrocal2010}. In contrast, \cite{Fuentes2005} address the change-of-support problem directly by expressing coarse-resolution output as the integral/average over high-resolution grid cells, and build models using observations as ground truth.

We also note that extensive work has been done on Gaussian process (GP) modeling for computer models from which output can be viewed as spatial (or spatio-temporal) data. However, the goals of computer model calibration \citep[e.g.,][]{Sacks1989, Kennedy2001}, are different from those of spatial downscaling in this paper. In computer model calibration, a primary interest lies in estimating context-specific inputs by building a statistical model for \textit{both} computer model outputs and physical observations \textit{together}, where coarse-resolution model outputs are treated as low-accuracy data and a discrepancy term is included to model the gap between computer models and physical reality \cite[e.g.,][]{Kennedy2000, Kennedy2001, Higdon2004}. Since OSSEs are typically used to evaluate new observing systems or data assimilation algorithms when actual observational data are \textit{not} available, our downscaling problem uses a \textit{single} data source, that is, our method uses the numerical model output \textit{solely}, \textit{without any} additional observations. As such, we focus here on developing a principled model-based framework to construct high-resolution NRs directly from coarse-resolution numerical model output.  

The problem presents the following challenges. (1) Atmospheric processes usually present nonstationary spatial structures. It is not realistic to model them with simple parametric spatial covariance functions such as the Mat\'ern family. More flexible covariance function models are required. (2) Although the numerical model output is obtained at coarse spatial resolution, the size of the numerical model output can still be large. For our application, we construct global NRs of atmospheric CO$_2$ concentration at high spatial resolution using output from the PCTM/GEOS-4/CASA-GFED atmospheric model, which is coupled with biospheric, biomass burning, oceanic, and anthropogenic $\text{CO}_2$ flux estimates \citep{Kawa2004, Kawa2010}. This model is referred to as PCTM hereafter. The PCTM output is generated at  $1^\circ \text{ latitude} \times 1.25^\circ  \text{ longitude}$, resulting in a dataset of size $M=52,128$. Such a large dataset will cause computational bottlenecks for traditional spatial statistics methods due to the computational cost of the Cholesky factorization for the associated large covariance matrix, and memory limitations. This is the well-known ``big $n$'' problem in spatial statistics \cite[e.g.,][]{Cressie2006, Cressie2008, Banerjee2008, Nychka2015}. (3) The difference between spatial resolutions of the numerical model output and the desired high-resolution NRs needs to be resolved carefully, and  taken into account in the downscaling procedure. Since the numerical model output is assumed to be the ``truth'' at its corresponding resolution, it is important that the resulting downscaled NRs maintain certain essential properties of the coarse-resolution numerical model output. Previous work addresses some of these issues. For example, \cite{Datta2015} propose a model to handle massive spatial data, focusing on the second issue in particular. \cite{Gramacy2015} provide a computationally efficient way to model large computer model outputs with nonstationary covariance, thus addressing both the first and second issues.  To our knowledge, all three issues have not been discussed within a unified general modeling strategy in previous work.

Our approach to spatial downscaling follows the classical additive model in geostatistics, with several components that characterize different spatial variabilities. As suggested in \cite{Fuentes2005} and \cite{Craigmile2009}, we address the change-of-support issue by building our statistical model for the spatial process at the highest resolution of interest. To infer the process at fine resolution from coarse-resolution data, we use a spatial process model imposed with necessary assumptions to avoid an ill-posed inverse problem. Such a strategy is referred to as geostatistical regularization in geostatistics \citep{Atkinson2001}. To alleviate computational difficulties associated with large data volumes, we use a nonstationary covariance function model that combines a low-rank component for dimension reduction and a component with a diagonal covariance matrix and/or sparse precision matrix. We extend Fixed Rank Kriging \citep{Cressie2008} and Fused Gaussian Process \citep{Ma2017} by making them more flexible. We capture nonstationary spatial variability with a forward stepwise algorithm for basis function selection. This includes both their locations and bandwidths of basis functions in the low-rank component. 

Our basis function selection method differs from that of \cite{Tzeng2017} in that our method is designed to learn nonstationary and localized features from the data. \cite{Tzeng2017} uses information about data locations, but not data values, to specify basis functions. Other methods for nonstationary spatial modeling such as \cite{Katzfuss2013} and \cite{Konomi2014} require computationally intensive reversible jump Markov chain Monte Carlo methods. In contrast, the forward stepwise algorithm we propose is simpler, more intuitive, and well-suited for parallel computing environments. The spatial downscaling procedure is also computationally efficient, and can produce not just one but many high-resolution statistical replicates from a coarse-resolution spatial field because it is based on conditional simulation.

Finally, the downscaled fields produced by our method maintain two important relationships with the coarse-resolution model output. First,  the spatial dependence structure of the downscaling model is estimated, and thus inherited, from the coarse-resolution data. Second, when aggregated back to the coarse resolution, our high-resolution NRs match the coarse-resolution data exactly. Note that the numerical model output is considered as the best representation of the geophysical process of interest and is used as the ``truth'' at the coarse resolution. Any departure from this output solely due to the downscaling process cannot be physically justified. Therefore, we impose this aggregation requirement when generating downscaled NRs. Such a requirement has been emphasized in various environmental studies. \cite{Zhou2009} show that it is important to explicitly resolve the discrepancy between the coarse and fine resolutions by accounting for the relationship between the known, aggregated observations and the unknown fine-resolution attributes. In climate science, a similar aggregation constraint (also called dynamic downscaling) is used in regional climate modeling \citep[e.g.,][]{Wilby1997}, when coarse-resolution outputs from global climate models are used as boundary conditions for regional climate models. Meanwhile, from the statistical perspective, this aggregation requirement stems directly from the change-of-support property, where spatial observations at coarse resolution are defined as stochastic integrals over the fine-resolution process.
 
The remainder of this paper is organized as follows. In Section~\ref{sec: methodology}, we formulate the spatial statistical model for downscaling and inference, including parameter estimation and downscaling, via conditional simulation. The forward basis function selection algorithm is also described. In Section~\ref{sec: simulation examples}, we present simulation studies to evaluate the performance of the proposed downscaling method and basis function selection algorithm. In Section~\ref{sec: application on pctm}, the methodology is applied to surface CO$_2$ concentrations produced by PCTM at $M=52,128$ grid cells to produce a high-resolution, downscaled field of $N=655,362$ equal-area hexagons over the globe. Section~\ref{sec: conclusion} concludes with discussion and future work.

\section{Methodology} \label{sec: methodology}
In this section, we present our model-based spatial downscaling framework. In particular, Section~\ref{sec: spatial model} introduces the spatial statistical model and Section~\ref{subsec:para_est} presents basic derivations for parameter estimation via the EM algorithm. The downscaling procedure via conditional simulation is given in Section~\ref{sec: conditional simulation}. Finally, Section~\ref{sec: adaptive basis algorithm} presents the forward basis function selection procedure.

\subsection{The Spatial Statistical Model} \label{sec: spatial model}

Let $\{ Y(\bs): \bs \in \mathcal{D}\}$ denote the atmospheric process of interest over a continuous spatial domain $\mathcal{D}\subset \mathbb{R}^d$ where $d\geq 1$ denotes the dimension of the spatial domain, and $\bs$ is a spatial location in $\mathcal{D}$. We consider the following additive model for the spatial process $Y(\cdot)$:
\begin{eqnarray} \label{eqn: data model}
Y(\bs) = \mu(\bs) + w(\bs) + \epsilon(\bs),\, \bs \in \mathcal{D},
\end{eqnarray}
where $\mu(\cdot)$ is a trend term incorporating important covariates. The second term in \eqref{eqn: data model}, $w(\cdot)$, is assumed to be a Gaussian process with mean zero, and covariance function $C(\bs_1,\bs_2)\equiv \text{cov}\{w(\bs_1), w(\bs_2)\}$ for $\bs_1,\,\bs_2\in\mathcal{D}$. The third term, $\epsilon(\cdot)$, is modeled as a Gaussian white-noise process in space with mean zero and variance $\sigma^2_{\epsilon}>0$, independent of $w(\cdot)$.

With a large number of observations, parameter estimation and prediction in kriging and Gaussian process regression become computationally infeasible. Many methods have been proposed to address this problem: covariance tapering \citep{Furrer2006}, composite likelihoods \citep{Eidsvik2014}, Gaussian Markov random fields \citep{Lindgren2011}  using low- and/or high-dimensional random vectors to induce covariance structures that result in low-rank covariance matrices and/or sparse precision matrices \cite[e.g.,][]{Banerjee2008, Cressie2008, Sang2012, Nychka2015, Datta2015, Katzfuss2016}, and local kriging \cite[e.g.,][]{Hammerling2012, Gramacy2015, Tadic2015}. Most of these methods rely on the assumption that $C(\bs_1,\bs_2)$ is stationary and/or has a prespecified parametric form, such as the Mat\'ern covariance family. \cite{Cressie2008} and \cite{Ma2017} take a different semiparametric approach. They use spatial basis functions, and allow the form of the covariance function to be flexible. We use such an approach for two reasons. First, it provides a globally valid spatial process model over the domain for \textit{joint} inference at \textit{all} locations. Second, it provides increased flexibility for modeling nonstationary behavior over a large spatial domain, as is typically the case for atmospheric processes. 

We assume that the process $w(\cdot)$ is induced by two independent components:
\begin{eqnarray} \label{eqn: w}
w(\bs) = \nu(\bs) + \delta(\bs) ,\, \bs \in \mathcal{D},
\end{eqnarray}
where the first term $\nu(\cdot)$ has a basis function representation: $\nu(\bs) = \bS(\bs)^T \boeta,\, \bs \in \mathcal{D}$. Here,  $\bS(\cdot)=(S_1(\cdot), \ldots, S_r(\cdot))^T$ is a vector of $r$ basis functions where $r$ is relatively small, and so we call this component $\nu(\cdot)$ the \textit{low-rank} component. The $r$-dimensional random vector $\boeta$ is assumed to follow the multivariate normal distribution with mean zero and covariance matrix $\bK$. We further assume the $r\times r$ covariance matrix $\bK$ to be a general symmetric positive definite matrix without any pre-specified form, allowing for great flexibility in modeling spatial dependence structure. The basis functions are chosen to be compactly-supported. We describe a data-driven approach to automatically select the centers and bandwidths of these basis functions in Section~\ref{sec: adaptive basis algorithm}. We will illustrate the procedure in Sections~\ref{sec: simulation examples} and \ref{sec: application on pctm}. In particular, we will show that the choice of basis functions can have a substantial impact on inference, especially predictive performance.

For the second term in \eqref{eqn: w}, we assume that the process $\delta(\cdot)$ is induced by a high-dimensional random vector $\bxi$: $\delta(\bs) = \bB(\bs)^T \bxi,\, \bs \in \mathcal{D}$. The vector $\bxi$ is defined by a discretization of the spatial domain $\mathcal{D}$ into a fine-resolution lattice (which can be irregular) of $N$ grid cells $\mathcal{D}\equiv \cup \{\bs_i\in \mathcal{A}_i: i=1, \ldots, N\}$ with $\{\mathcal{A}_i: i=1, \ldots, N\}$ called basic areal units (BAUs), as suggested in \cite{Nguyen2012}.  In practice, these BAUs are determined by the finest resolutions of interest that are required to construct synthetic observations and forecasts in OSSEs. Note that $N$ can be very large, and can be much larger than the number of observed data points. We further model $\bfxi$ with a Gaussian random Markov field, particularly, the spatial conditional autoregressive (CAR) model.  The precision matrix of $\bfxi$ is assumed to be: $\mathbf{Q}\equiv (\mathbf{I} - \gamma \mathbf{H})/\tau^2$, which is induced by the full conditional distributions $\xi_i|\{\xi_j: j\neq i\}\sim N(\gamma \sum_{j=1}^N H_{ij}\xi_j, \tau^2)$, for $i=\ldots, N$. Here, $\gamma$ is called the spatial dependence parameter. If $\gamma=0$, the elements in $\bfxi$ will be independent. The parameter $\tau^2$ is called the conditional marginal variance. The matrix $\mathbf{H}\equiv(H_{ij})_{i,j=1, \ldots, N}$ is an  $N\times N$ proximity matrix with $H_{ii}=0$, and $H_{ij}=1$ if $\mathcal{A}_j$ is a neighbor of $\mathcal{A}_i$ and is zero otherwise, where $H_{ij}=0$ for $i\neq j$ implies that $\xi_i$ and $\xi_j$ are \textit{conditionally} independent given $\{\xi_{\ell}: \ell\neq i, j\}$. To specify the neighborhood structure, one can choose a threshold distance in terms of spatial adjacency; see Chapter 6 of \cite{Cressie1993} for more details on specification of a CAR model and examples of neighborhood structures. Following \cite{Ma2017}, we choose the basis function vector $\bB(\cdot)\equiv (B_1(\cdot), \ldots, B_{N}(\cdot))^T$ to be a vector of incidence functions, where $B_i(\bs)=1$ if the location $\bs$ is in $\mathcal{A}_i$ and zero otherwise, for $i=1,\ldots, N$. For more complicated precision structures, other types of basis functions such as piecewise linear basis functions and Wendland basis functions can be used for $B_i(\cdot)$; for details see \cite{Ma2017}.

The model for the process $Y(\cdot)$ is thus,
\begin{eqnarray} \label{eqn: fgp}
Y(\bs)= \mu(\bs)+ \bS(\bs)^T\boeta +\bB(\bs)^T\bxi + \epsilon(\bs),
\end{eqnarray}
with covariance function
\begin{eqnarray} \label{eqn: fgp_cov}
C_{FGP}(\bs_1,\bs_2)\equiv \cov\{Y(\bs_1), Y(\bs_2)\}= \bS(\bs_1)^T \bK \bS(\bs_2) + \bB(\bs_1)^T \bQ^{-1}\bB(\bs_2) + \sigma^2_\epsilon I(\bs_1=\bs_2).
\end{eqnarray}
We call this model the Fused Gaussian Process model, referred to simply as FGP in the discussion that follows. Previous work in \cite{Cressie2010high} and \cite{Nguyen2012} assume that the process $\delta(\cdot)$ is a spatial white noise process with mean zero and variance $\sigma^2_{\xi}$. This model is called the spatial random effects model, and the resulting method is referred to as Fixed Rank Kriging (FRK) hereafter. The corresponding covariance function is given by,
\begin{eqnarray} \label{eqn: frk_cov}
C_{FRK}(\bs_1,\bs_2)\equiv \cov\{Y(\bs_1), Y(\bs_2)\}= \bS(\bs_1)^T \bK \bS(\bs_2) + \sigma^2_\xi I(\bs_1=\bs_2) + \sigma^2_\epsilon I(\bs_1=\bs_2).
\end{eqnarray} 

In the case studied here, rather than observing the process $Y(\cdot)$ at BAU-level, we only have the aggregated values of this process at coarse spatial resolution, i.e., the coarse-resolution numerical output. The difference between the spatial resolution of this numerical model output and the NRs we need is a type of change-of-support problem \cite[e.g.,][]{Cressie1993, Gotway2002, Wakefield2017}. Suppose that the numerical model output is obtained over a total of $M$ coarse grid cells, $\{\Delta_i \subset \mathcal{D}: i=1,\ldots,M\}$ in the spatial domain. We call the region $\Delta$ the \emph{support} of $Y(\Delta)$ and define $Y(\Delta)$ as the average of $Y(\cdot)$ over its support:
\begin{eqnarray} \label{eqn: cos}
Y(\Delta) := \frac{1}{|\Delta|} \int_{\bs \in \Delta} Y(\bs)\, d\bs, 
\end{eqnarray}
where $|\Delta|>0$ is the volume of $\Delta$. This stochastic integral is defined as a mean-square limit, which can be approximated by an appropriately weighted sum \cite[see][Section 5.2]{Cressie1993}.  An alternative and more flexible definition is $Y(\Delta)\equiv \int_{\bs\in\Delta} Y(\bs)h(\bs)d\bs$, where $h(\bs)$ is called the impulse response or the point spread function in remote sensing science. Eq.~\eqref{eqn: cos} assumes additionally that $h(\bs)=\frac{1}{|\Delta|}$, if $\bs\in \Delta$, and 0, otherwise. It is also possible to use a non-constant impulse response \citep[e.g.,][]{Cracknell1998}. Although we focus on the constant case \eqref{eqn: cos} in this work, it is straightforward to apply our method and algorithms with a more general impulse response. 

 To relate  $\Delta$ to fine-resolution BAUs, the integral \eqref{eqn: cos} is approximated by 
\begin{eqnarray} \label{eqn: approximate cos}
Y(\Delta) \approx \frac{1}{\sum_{\bs \in \mathcal{D}} I(\bs \in \Delta)}  \sum_{\bs \in \mathcal{D}} I(\bs \in \Delta) \cdot Y(\bs),
\end{eqnarray}
where the summation is taken over the discretized domain $\mathcal{D}$ with $N$ BAUs, and $I(\bs \in \Delta)$ is an indicator function that is equal to one if the centroid $\bs$ of $\mathcal{A}$ lies in the region $\Delta$, and is equal to zero otherwise. 

Let $\widetilde{\bY}\equiv (Y(\Delta_1), \ldots, Y(\Delta_M))^T$ be a vector of the numerical model output for $M$ coarse-resolution grid cells. We are interested in recovering the process $Y(\cdot)$ at BAU-level,  $\bY\equiv (Y(\bs_1), \ldots, Y(\bs_N))^T$, from $\widetilde{\bY}$. We define the so-called $M\times N$ aggregation matrix $\bA$ whose $(i,j)$-th entry $a_{ij}$ is given by,
\begin{eqnarray} \label{eqn: aggregation matrix}
a_{ij} \equiv \frac{I(\bs_j \in \Delta_i)}{\sum_{\bs \in \mathcal{D}} I(\bs \in \Delta_i)},\, i=1,\ldots, M; j=1, \ldots, N.
\end{eqnarray}
Recall that the FGP and FRK models are defined at the BAU resolution. Fortunately, it is straightforward to obtain the \textit{marginal} distributions of $\bY$ and $\widetilde{\bY}$: $\bY\sim \mathcal{N}_N(\bmu, \bfSigma)$ 
and $\widetilde{\bY}\sim \mathcal{N}_M(\bA \bmu,  \bA\bfSigma\bA^T)$, where $\bmu=(\mu(\bs_1), \ldots, \mu(\bs_N))^T$ is a vector of trends terms defined at BAU-level. The covariance matrix of $\bY$ can be easily derived from the covariance functions given in Eq.~\eqref{eqn: fgp_cov} and Eq.~\eqref{eqn: frk_cov}: 
\begin{eqnarray} \label{eqn: covariance matrix at BAU}
\bSigma \equiv \text{cov}(\bY) =  \bS\bK \bS^T + \bSigma_{\delta} + \bV,
\end{eqnarray}
where $\bS$ is the $N\times r$ matrix with its $i$-th row defined as the transpose of $\bS(\bs_i)$ for $i=1,\ldots, N$. The $N\times N$ matrix $\bSigma_{\delta}$ is obtained from the process $\delta(\cdot)$, and takes the form $\bSigma_{\delta}=\bQ^{-1}$ in FGP and $\bSigma_{\delta}=\sigma^2_{\xi} \mathbf{I}_N $ in FRK. The last term $\bV\equiv \sigma^2_{\epsilon}\mathbf{I}_N$ is an $N\times N$ matrix resulting from the process $\epsilon(\cdot)$ in Eq.~\eqref{eqn: data model}. When the spatial dependence parameter $\gamma=0$, the covariance matrix from FGP is reduced to that of FRK. Under both models,  the number of basis functions, $r$, is assumed to be much smaller than the number of data points, $M$, which provides substantial dimension reduction.

Note that both FRK and FGP inherit an additive structure widely used in modeling spatial data. The modified predictive process \citep{Finley2009}, the full scale approximation \citep{Sang2012}, and the multi-resolution approximation \citep{Katzfuss2016}, all construct additive models based the assumption of a particular parametric covariance function such as the Mat\'ern covariance function. \cite{Ba2012} use a combination of two spatial covariance structures together, but this requires empirical constraints on parameters, and is not designed to handle large datasets. Comparing FRK and FGP,  the latter introduces spatial dependence for the term $\boldsymbol\xi$ so that the resulting model can give better predictive performance than typical low-rank models including FRK. Numerical examples to demonstrate the robust predictive performance of FGP for different covariance functions can be found in \cite{Ma2017}. They also discuss other assumptions on $\bfxi$, besides the CAR model for FGP.

\subsection{Parameter Estimation}\label{subsec:para_est}
The parameters of the models proposed in Section~\ref{sec: spatial model} are estimated using  likelihood-based approaches. We follow the approach of \cite{Cressie2010high} and \cite{Nguyen2012} and assume that the variance parameter of $\epsilon(\cdot)$, $\sigma^2_\epsilon$, is known from independent validation data or estimated separately by examining the empirical variograms \citep{Kang2010}. The trend term is assumed to be $\mu(\cdot)=\bX(\cdot)^T \bbeta$ with a $p$-dimensional vector of known covariates $\bX(\cdot) = (X_1(\cdot), \ldots, X_p(\cdot))^T$ and corresponding unknown coefficients $\bbeta$. Let $\btheta$ denote the set of parameters to be estimated. For FGP, $\btheta$ consists of $\{\bbeta, \bK, \tau^2, \gamma\}$, and for FRK, $\btheta\equiv \{\bbeta, \bK, \sigma^2_{\xi}\}$. 

Recall that the ``observed'' data come from the numerical model output, $ \widetilde \bY$, which is assumed to follow the multivariate normal distribution with mean $E(\widetilde \bY) = \bA \bmu$, and covariance matrix $\text{cov}(\widetilde \bY) = \bA \bSigma \bA^T$. Up to an additive constant, the corresponding twice-negative-marginal-log-likelihood function is,
\begin{eqnarray} \label{eqn: likelihood fun}
-2\ln L(\btheta | \widetilde \bY) = \ln | \bA \bSigma \bA^T| + (\widetilde \bY - \bA \bX\bbeta)^T(\bA \bSigma \bA^T)^{-1}(\widetilde \bY - \bA \bX\bbeta),
\end{eqnarray}
where $\bX$ is the $N\times p$ design matrix associated with the covariates, and the covariance matrix $\bSigma$ is given in Eq.~\eqref{eqn: covariance matrix at BAU}. Note that evaluating Eq.~\eqref{eqn: likelihood fun} requires inverting and calculating the log-determinant of the $M\times M$ matrix $\bA\bSigma \bA^T$. Specifically, we have
\begin{eqnarray} \label{eqn: ASigmaA}
\bA\bSigma\bA^T=(\bA\bS)\bK (\bA\bS)^T +\bD^{-1},\text{ where } \left\{
\begin{array}{l}
\bD=[\bA (\sigma^2_{\xi} \mathbf{I}) \bA^T +  \sigma^2_{\epsilon}\bA\bA^T]^{-1}\text{ for FRK;}\\
\bD=(\bA \bQ^{-1} \bA^T +  \sigma^2_{\epsilon}\bA\bA^T)^{-1}\text{ for FGP.}\\
\end{array}
\right.
\end{eqnarray}

Although the numerical model output, $\widetilde \bY$, is defined at coarse spatial resolution, the resulting number of grid cells, $M$, can still be large. For example, in the application of downscaling atmospheric CO$_2$ concentrations in Section~\ref{sec: application on pctm}, $M=52,128$. Recall that in the definition of $\bA$ in \eqref{eqn: aggregation matrix}, the $(i,j)$-th element in the product of $\bA$ and $\bA^T$ is given by
 $$(\bA\bA^T)_{ij}=\sum_{k=1}^{ N }a_{ik}a_{jk}; \,\,i,j=1,\ldots,M.$$
Since any BAU $\mathcal{A}$ is assumed to be uniquely associated with a single coarse-resolution grid cell, at least one of $a_{ik}$ and $a_{jk}$ is zero whenever $i\neq j$, and the $M\times M$ matrix $\bA\bA^T$ is diagonal. For FRK, the matrix $\bD=[\bA (\sigma^2_{\xi}\mathbf{I}) \bA^T + \bA (\sigma^2_{\epsilon}\mathbf{I})\bA^T]^{-1}$ is diagonal as well. Thus, the matrix $\bA\bSigma \bA^T$ can be inverted by applying the Sherman-Woodbury-Morrison formula \cite[e.g.,][]{Hendeson1981} as follows:
\begin{eqnarray} \label{eqn: inverse of cov matrix}
(\bA\bSigma\bA^T)^{-1} = \bD - \bD(\bA\bS)[\bK^{-1} + (\bA\bS)^T \bD(\bA\bS)]^{-1} (\bA\bS)^T \bD.
\end{eqnarray}
This only requires inverting diagonal and low-rank ($r\times r$) matrices. To calculate the determinant $|\bA\bSigma\bA^T|$ for FRK, we use Sylvester's determinant identity \cite[see][]{Akritas1996}:\begin{eqnarray} \label{eqn: determinant}
|\bA \bSigma \bA^T| &=& |(\bA\bS) \bK (\bA\bS)^T + \bD^{-1}| = |\bK^{-1} + (\bA \bS)^T\bD (\bA \bS)| |\bK| | \bD^{-1}|,
\end{eqnarray}
which involves determinants of diagonal and $r\times r$ matrices. For FGP, $\bD=(\bA \bQ^{-1} \bA^T +  \sigma^2_{\epsilon}\bA\bA^T)^{-1}$, where $\bQ$ is a sparse matrix. The Sherman-Morrison-Woodbury formula can be used to calculate $\bD$ in \eqref{eqn: inverse of cov matrix} as well,
\begin{eqnarray} \label{eqn: D_inv_FGP}
\bD=(\sigma^2_{\epsilon} \bA\bA^T)^{-1} - (\sigma^2_{\epsilon} \bA\bA^T)^{-1} \bA [\bQ + \bA^T (\sigma^2_{\epsilon} \bA\bA^T)^{-1} \bA]^{-1} \bA^T (\sigma^2_{\epsilon} \bA\bA^T)^{-1}.
\end{eqnarray}
This only requires solving a sparse linear system. To calculate $|\bD^{-1}|$ in \eqref{eqn: determinant} for FGP, Sylvester's determinant identity can be used again: 
\begin{eqnarray} \label{eqn: determinant_FGP}
|\bD^{-1}|=|\bQ+\bA^T(\sigma^2_{\epsilon}\bA \bA^T)^{-1} \bA| | \bQ^{-1}| | \sigma^2_{\epsilon}\bA \bA^T|.
\end{eqnarray}
So, in both FRK and FGP, the twice-negative-marginal-log-likelihood function in~\eqref{eqn: likelihood fun} can be computed efficiently.

Here, we adapt the EM algorithms used in \cite{Katzfuss2011} and \cite{Ma2017} to obtain maximum-likelihood estimates of the parameters $\btheta$ using data $\widetilde\bY$. Specifically, the random vector $\boeta$ is treated as ``missing data'', and the ``complete data" likelihood $L_C(\boeta, \widetilde \bY)$ can be obtained. Up to an additive constant, the twice-negative-complete-data-log-likelihood function is given by 
\begin{eqnarray} \label{eqn: complete data log-likelihood}
-2\ln L_C(\boeta, \widetilde \bY) &= &\ln | \bD^{-1} | + [\widetilde \bY - (\bA\bX)\bbeta-\bA\bS \boeta]^T \bD [\widetilde \bY - (\bA\bX)\bbeta-\bA\bS \boeta] \\ \nonumber
&&+ \ln | \bK| + \boeta^T \bK^{-1} \boeta.
\end{eqnarray}
In the E-step of the EM algorithm, the conditional distribution of $\boeta$ given $\widetilde \bY$ under parameters $\btheta$ is multivariate normal with mean $\bmu_{\boeta| \widetilde\bY, \btheta} = \bK (\bA \bS)^T (\bA \bSigma \bA^T)^{-1} (\widetilde \bY-\bA\bX\bbeta)$ and covariance matrix $\bSigma_{\boeta| \widetilde\bY, \btheta} = \bK - \bK(\bA \bS)^T (\bA \bSigma \bA^T)^{-1} (\bA \bS) \bK^T$. Then the conditional expectation of $\ln L_C(\boeta, \widetilde \bY)$ with respect to the distribution $[\boeta| \widetilde \bY]$, referred to as the $Q$ function, can be derived. In the M-step, parameters are updated by finding the maximum of this $Q$ function with respect to $\btheta$. For FRK, all parameters in $\btheta$ have closed-form updates, while in FPG numerical optimization algorithms are used to update $\tau^2$ and $\gamma$. Note that the results in \eqref{eqn: inverse of cov matrix}, \eqref{eqn: determinant}, \eqref{eqn: D_inv_FGP}, and \eqref{eqn: determinant_FGP} allow efficient computation in the execution of the EM algorithm. Details of the parameter estimation scheme via the EM algorithm are given in Appendix~\ref{appendix: EM}.

\subsection{Statistical Downscaling via Conditional Simulation} \label{sec: conditional simulation}

Recall that numerical model output $\widetilde{\bY}$ is defined at coarse spatial resolution over grid cells $\{\Delta_i: i=1, \ldots, M\}$. In Section~\ref{sec: spatial model}, we gave the resulting distribution of $\widetilde{\bY}$ under FRK and FGP, respectively. Our goal is to simulate the process $Y(\cdot)$ at fine-resolution, i.e., over all BAUs $\{\mathcal{A}_i: i=1, \ldots,N\}$, given the numerical model output. To avoid introducing additional notation, we also use $\widetilde{\bY}$ to represent a realization of this random vector: the observed numerical model output. In OSSEs, numerical models represent state-of-the-art understanding of atmospheric processes, and so their output is assumed to be the ``truth'' at its corresponding resolution. Therefore, when we downscale $\widetilde{\bY}$ to construct NRs at the resolution of the BAUs, we impose the hard constraint that fine-resolution NRs should match the numerical model output exactly when the NRs are aggregated from BAU-level back up to the coarse-resolution grid cells.

The conditional distribution of $\bY$ given $\widetilde \bY$ is derived as follows. To ensure that the simulated $\bY$ match $\widetilde{\bY}$ after aggregation, we impose the constraint that $\bA\bY=\widetilde{\bY}$. Using the standard result for conditional distributions of multivariate normal distributions \citep[e.g.,][pp. 156-157]{Ravishanker2002}, the conditional distribution of $\bY$ given $\widetilde\bY$ is:
\begin{eqnarray} \label{eqn: conditional distribution}
\bY\mid \bA\bY=\widetilde \bY & \sim & \mathcal{N}_N(\bmu + \bSigma \bA^T (\bA \bSigma \bA^T)^{-1} (\widetilde{\bY} - \bA \bmu),\, \bSigma - \bSigma\bA^T (\bA \bSigma \bA^T)^{-1} \bA \bSigma).
\end{eqnarray}
To efficiently compute the conditional mean vector for FRK and FGP, we use the results in Eq.~\eqref{eqn: inverse of cov matrix} and Eq.~\eqref{eqn: D_inv_FGP} to evaluate $(\bA\bSigma\bA^T)^{-1}$. The conditional mean vector, $\bmu_{\bY|\widetilde\bY}\equiv\bmu + \bSigma \bA^T (\bA \bSigma \bA^T)^{-1} (\widetilde{\bY} - \bA \bmu)$, gives the optimal spatial predictions of $Y(\cdot)$ at BAU-level, given data $\widetilde\bY$, under squared-error loss. The associated prediction uncertainties,  i.e., the prediction variances, are the corresponding \textit{diagonal} elements in the conditional covariance matrix $\bSigma_{\bY|\widetilde{\bY}}\equiv\bSigma - \bSigma\bA^T (\bA \bSigma \bA^T)^{-1} \bA \bSigma$, and can also be calculated efficiently using Eq.~\eqref{eqn: inverse of cov matrix} and Eq.~\eqref{eqn: D_inv_FGP}.

To construct an ensemble of NRs we simply draw samples from the conditional distribution of $\bY$ given $\widetilde\bY$. Directly sampling from this conditional distribution in (\ref{eqn: conditional distribution}) requires storing the $N\times N$ covariance matrix $\bSigma_{\bY|\widetilde{\bY}}$ and  performing a Cholesky decomposition on it, which results in $O(N^2)$ memory cost and $O(N^3)$ flops. Note that with BAUs defined at fine-resolution, $N$ is very large. For example, in the application of downscaling surface CO$_2$ concentrations in Section~\ref{sec: application on pctm}, $N= 655,362$. Therefore, directly sampling from the distribution in (\ref{eqn: conditional distribution}) is prohibitive. To circumvent this problem, we devised a step-by-step procedure to generate a sample, denoted by $\bY_{\text{CS}}$, from the conditional distribution. The procedure is given by Algorithm~\ref{algorithm:cs}. The resulting random vector, $\bY_{\text{CS}}$, has some desirable properties, given in Proposition~\ref{Property of conditional simulation}. See Appendix~\ref{appendix: proof} for the proof.

\begin{theorem} \label{Property of conditional simulation}
The conditional sample $\bY_{\text{CS}}$ generated via Algorithm~\ref{algorithm:cs} has the following properties: 
\begin{itemize}[noitemsep,topsep=4pt]  
\item[(1)] The sample $\bY_{\text{CS}}$ satisfies the hard constraint: $\bA\bY_{\text{CS}}=\widetilde\bY$.
\item[(2)] The conditional distribution of $\bY_{\text{CS}}$ given $\bA\bY$ is multivariate normal with mean $E(\bY_{\text{CS}}\mid \bA\bY)= \bmu+\bSigma\bA^T(\bA\bSigma\bA^T)^{-1}(\bA\bY-\bA\bmu)$ and covariance matrix $\text{cov}(\bY_{\text{CS}}\mid\bA\bY)=\bSigma-\bSigma\bA^T(\bA\bSigma\bA^T)^{-1}\bA\bSigma$. Thus, given $\bA\bY=\widetilde\bY$, the random vector $\bY_{\text{CS}}$ follows the same distribution as $\bY$ given in Eq. \eqref{eqn: conditional distribution}.
\item[(3)] The marginal distribution of $\bY_{\text{CS}}$ is multivariate normal with mean $\bmu$ and covariance matrix $\bSigma$.
\item[(4)] With the parameters $\btheta$ known, if we define the mean-squared-error of $\bY_{\text{CS}}$ to be $E[\bY_{\text{CS}} - \bY]^2$,  then $E[\bY_{\text{CS}} - \bY]^2=2[\bSigma - \bSigma \bA^T (\bA\bSigma \bA^T)^{-1} \bA \bSigma]$.
\end{itemize}
\end{theorem}

\begin{algorithm}
	\caption{Generate a sample, $\bY_{\text{CS}}$, from the conditional distribution of $\bY$ given the $\bA\bY=\widetilde\bY$ in (\ref{eqn: conditional distribution}). Here, the subscript ``CS'' stands for conditional simulation.}
	\label{algorithm:cs}
	\begin{algorithmic}[1]
	 
	 \State {\bf Input:} The numerical model output $\widetilde \bY$, the $M\times N$ aggregation matrix $\bA$, $\sigma^2_\epsilon$, and  $\btheta$. For FGP, $\btheta$ consists of $\{\bbeta, \bK, \tau^2, \gamma\}$; for FRK, $\btheta\equiv \{\bbeta, \bK, \sigma^2_{\xi}\}$. 
	 
	 \noindent / / \textit{Generate $\bY_{\text{NS}}$, a sample from the marginal distribution of $\bY$. Here, the subscript ``NS'' stands for marginal or unconditional simulation.}
	 	 \State Generate a sample $\boeta_{\text{NS}}$ from $\mathcal{N}_r(\bfzero, \bK)$, requiring Cholesky decomposition of the $r\times r$ matrix $\bK$.
		 	 \State Generate a sample $\bepsilon_{\text{NS}}$ from $\mathcal{N}_N(\bfzero, \sigma^2_\epsilon\bI)$, i.i.d. normal random variables with mean zero and variance $\sigma^2_\epsilon$.
	 \State Generate a sample $\bdelta_{\text{NS}}$ from $\mathcal{N}_N(\bfzero, \bSigma_\delta)$:
	 	\begin{itemize}
		\item  For FRK, $\bSigma_\delta=\sigma^2_\xi\bI$, thus sampling $\bdelta_{\text{NS}}$ as i.i.d. random variables with mean zero and variance $\sigma^2_\xi$.
		\item For FGP, $\bSigma_\delta=\bQ^{-1}$; this sampling step requires Cholesky decomposition of the sparse matrix $\bQ\equiv(\bI-\gamma\bH)/\tau^2$.
		\end{itemize}
		\State  \textbf{Return} $\bY_{\text{NS}}=\bX\bbeta+\bS\boeta_{\text{NS}}+\bdelta_{\text{NS}}+\bepsilon_{\text{NS}}$, a sample from the marginal distribution. 
		 
		 \noindent/ / \textit{Adjust $\bY_{\text{NS}}$ to obtain $\bY_{\text{CS}}$, a sample from the  distribution in \eqref{eqn: conditional distribution} conditional on $\bA\bY=\widetilde\bY$.}
	\State Calculate $(\bA\bSigma\bA^T )^{-1}(\bA\bY-\bA\bY_{\text{NS}})$ with $\bA\bSigma\bA^T$ given in Eq.~\eqref{eqn: ASigmaA}: 
		\begin{itemize}
		\item For FRK, use the Sherman-Woodbury-Morrison formula as in Eq.~\eqref{eqn: inverse of cov matrix}.
		\item For FGP, use both Eq.~\eqref{eqn: inverse of cov matrix} and Eq.~\eqref{eqn: D_inv_FGP}.
		\end{itemize}
	\State \textbf{Return} $\bY_{\text{CS}} = \bY_{\text{NS}} + \bSigma \bA^T (\bA \bSigma \bA^T)^{-1} (\bA{\bY} - \bA \bY_{\text{NS}})$. 
	
	 \noindent  / / \textit{If more than one sample is needed, repeat Step 2 through Step 7, noting that calculating Cholesky decompositions of matrices only needs to be done once.}
	\end{algorithmic}
\end{algorithm}

High-resolution NRs can be constructed efficiently by drawing samples from the conditional distribution of $\bY$ given $\widetilde\bY$ using Algorithm~\ref{algorithm:cs}. As stated in Proposition 1,  the constraint $\textbf{A}\bY_{\text{CS}} = \widetilde{\textbf{Y}}$ is satisfied, implying that when the high-resolution NRs are aggregated back to  coarse resolution, exactly match the numerical model output, $\widetilde\bY$. In addition, the high-resolution NR, $\bY_{\text{CS}}$, has a marginal distribution based on the models defined for the process $Y(\cdot)$ at the finest scale as discussed in Section~\ref{sec: spatial model}. Parameters are assumed known in Algorithm~\ref{algorithm:cs}. In practice they are estimated from coarse-resolution output, $\widetilde\bY$, using the EM algorithm as described in Section~\ref{subsec:para_est}. 

We conclude this section with remarks about computational complexity related to the downscaling procedures for the two models FRK and FGP in Section~\ref{sec: spatial model}. As shown in \cite{Cressie2008} and \cite{Ma2017}, both FRK and FGP have desirable change-of-support properties, since the integration in Eq.~\eqref{eqn: cos} and summation in Eq.~\eqref{eqn: approximate cos} for basis functions can be done offline. With sparse matrices $\bA$ and $\bS$, the product of $\bA$ and $\bS$ can be computed efficiently with, at most, $O(Mrb_0)$ flops for both FRK and FGP, where $b_0$ is the maximum number of BAUs falling into a single coarse-resolution grid cell. To generate a unconditional sample $\bY_{\text{NS}}$ from its marginal distribution, FRK requires $O(Nr+r^3)$ flops, while FGP requires $O(Nr + N^{1.5})$ flops. To adjust $\bY_{\text{NS}}$ to obtain $\bY_{\text{CS}}$, evaluation of $(\bA\bSigma\bA^T)^{-1} \mathbf{b}$ for a vector $\mathbf{b}$ of length $M$ is needed. Solving $(\bA\bSigma\bA^T)^{-1} \mathbf{b}$ requires $O(Mr^2)$ flops for FRK, since it only needs to calculate inversions of $r\times r$ matrices and $n\times n$ diagonal matrices, as well as multiplication of $M\times r$ and $r\times r$ matrices. Therefore, the overall computational cost is $O(Mrb_0+Mr^2+Nr)$ for spatial downscaling based on FRK. For FGP, solving $(\bA\bSigma\bA^T)^{-1} \mathbf{b}$ requires Cholesky decomposition of sparse matrix $\bQ$ and $\bQ+\bA^T(\sigma^2_{\epsilon}\bA\bA^T)^{-1}\bA$. As discussed in \cite{Rue2005}, the Cholesky decomposition of $\bQ$ has $O(N^{1.5})$ computational cost for a two-dimensional domain. Notice that the matrix $\bA^T(\sigma^2_{\epsilon}\bA\bA^T)^{-1}\bA$ is a block diagonal matrix with at most $b_0^2$ nonzero elements for each of $\floor*{N/b_0}$ block matrices. Hence, the sparse matrix $\bQ+\bA^T(\sigma^2_{\epsilon}\bA\bA^T)^{-1}\bA$ has $2Nb_0$ nonzero elements at most given the fact that the number of nonzero elements in $\bQ$ is smaller than $Nb_0$. The Cholesky decomposition of $\bQ+\bA^T(\sigma^2_{\epsilon}\bA\bA^T)^{-1}\bA$ has computational cost $O(N(p_0^2+3p_0))$ after appropriate reordering to obtain a band matrix with its bandwidth $p_0\ll N$, though solving the sparse linear system associated with matrix $\bQ+\bA^T(\sigma^2_{\epsilon}\bA\bA^T)^{-1}\bA$ can cost more computationally than solving that associated with $\bQ$. Therefore, the overall computational cost is $O(Mrb_0+2N(r^2+p_0^2)+N^{1.5}r)$ at most for spatial downscaling based on FGP. Note that in practice, the fine-resolution grid and corresponding BAUs are available based on the scientific goals of the observing systems or data assimilation algorithms. Because its computational complexity is related to $N$, it is neither practically necessary nor computationally economical to define the spatial process at a finer spatial resolution than what is needed in OSSEs. For the memory cost, both FRK and FGP require storage of sparse matrices $\bA$, $\bS$, $\bA \bS$, and the diagonal matrix $\bA \bA^T$, which have $O(Nr)$ memory cost at most. FGP also requires storage of the Cholesky factor of an $N\times N$ sparse matrix and $N\times r$ sparse matrix, which has $O(N\log N + Nr)$ memory cost at most. As $\log N \ll r$, $O(N\log N)$ is upper bounded by $O(Nr)$. Therefore, the overall memory cost is $O(Nr)$ in both FRK and FGP.

\subsection{The Forward Basis Function Selection Algorithm} \label{sec: adaptive basis algorithm}
Basis function selection has been investigated for low-rank methods used in analyzing large spatial data sets. Many methods \cite[e.g.,][]{Banerjee2008, Sang2012, Katzfuss2016} require pre-specification of a parametric covariance function that, in practice, is usually chosen to be  stationary. For these methods, basis functions are determined by the locations of knots, and a pre-specified parametric covariance function. 
\cite{Finley2009} propose an algorithm to sequentially find such  knot locations. For semiparametric methods including FRK and FGP, while they avoid the assumption of a specific parametric covariance function, basis functions do need to be specified. \cite{Cressie2008} recommend compactly-supported multiresolution functions, where the centers and bandwidths of basis functions for each resolution need to be specified. Specifically, a fixed number of resolutions (typically two or three) is chosen first. Then, for the $i$-th resolution, users specify the total number of basis functions, say, $r_i$. The centers of these $r_i$ basis functions are then regularly placed over the spatial domain, and all use the same bandwidth. We call such basis functions \emph{equally-spaced} basis functions hereafter. Readers are referred to \cite{Cressie2008} and \cite{Nguyen2012, Nguyen2014} for examples of these equally-spaced basis functions. \cite{Zhu2015} and \cite{Zammit2017} discuss the use of these equally-spaced basis functions, and suggest removing those where data are rare in order to achieve stable estimation. To the best of our knowledge, no method has been suggested to select both the centers and bandwidths of basis functions in a data-driven way. 

Our method sequentially adds new basis function centers and specifies their bandwidths based on their potential ability to improve spatial prediction. This protocol directly addresses one of the primary purposes of spatial data analysis, i.e., spatial prediction, and tends to perform very well in capturing nonstationary spatial variability. See our simulation study in Section~\ref{sec: toy example for basis selection algorithm}. For the numerical examples in this paper, we focus on one particular form of Wendland basis functions \citep{Wendland1995}: $S(\mathbf{u})=(1-\|\mathbf{u}-\mathbf{c}\|/r)^4I(\|\mathbf{u}-\mathbf{c}\|\leq r),\, \mathbf{u}\in \mathcal{D}$, where $\mathbf{c}$ is the center and $r$ is the bandwidth. This form of the Wendland basis function belongs to the family of compactly-supported basis functions. Compactly-supported basis functions have been widely adopted in previous studies \cite[e.g.,][]{Cressie2006, Cressie2008, Nguyen2012, Nguyen2014, Nychka2015, Shi2017, Ma2017}. For example,  \cite{Cressie2008} use the bisquare basis functions, and \cite{Nychka2015} use the Wendland basis functions that are different from ours. \cite{Zammit2017} discuss other types, such as Gaussian basis functions, that also require specification of center and bandwidth. The method we propose can be used for other types of basis functions as well.

The basic idea is as follows. First, we define a finite set of $r_*$ locations spread out across the entire domain $\mathcal{D}$ of interest. This set is referred to as a set of \emph{candidate centers}, denoted by $\mathcal{S}^*\equiv\{ \mathbf{s}_i \in \mathcal{D}: i=1, \ldots, r_*\}$. Suppose that we pre-specify a set of $r^{(1)}$ basis functions with centers $\mathcal{C}^{(1)}\equiv\{\bc_{1,1}, \ldots, \bc_{1,r^{(1)}}\}$ and bandwidths $\mathcal{B}^{(1)}\equiv\{b_{1,1}, \ldots, b_{1, r^{(1)}}\}$ as the initial sets of centers and corresponding bandwidths, where the superscript stands for the iteration of the algorithm. A possible choice is a small set of equally-spaced basis functions over the domain. The centers and bandwidths of new basis functions are added automatically at each iteration of the forward algorithm. At the beginning of the $i$-th iteration ($i\geq 1$), the current set of basis functions is used to fit data with the FRK model, which gives the estimated trend $\hat\mu(\cdot)$ and the estimated random effects, $\hat\boeta$. We use the pseudo-residuals defined as $D^{(i)}(\bs) \equiv Y(\bs) - \hat\mu(\bs) - \bS^{(i)}(\bs) \hat\boeta$, where $Y(\bs)$ is the observation at location $\bs$ and $\bS^{(i)}$ denotes the matrix resulting from the current set of basis functions, to assess where basis functions should be added to improve the fit. As in classical geostatistics \citep{Cressie1993}, these pseudo-residuals can be used to carry out the local semivariogram analysis for each observation location. The empirical-local-mean-squared error (ELMSE) is defined for each point in the candidate centers $\mathcal{S}^*$: $L^{(i)}(\bs)\equiv \var\{{D}^{(i)}(\bu): \bu\in \mathcal{N}(\bs)\} + (\text{ave}\{{D}^{(i)}(\bu): \bu\in \mathcal{N}(\bs)\})^2$, where $\mathcal{N}(\bs)$ denotes a local neighborhood surrounding the location $\bs$ and is chosen based on the effective range obtained from the semivariogram analysis. The effective range is defined as the distance at which the semivariogram value achieves 95\% of the sill. New basis functions are placed where the ELMSE is large, and the bandwidths of these basis functions are chosen corresponding to the effective range. We also impose a separation criterion to avoid substantial overlap among the supports of the newly-added basis functions at each iteration. We repeat these steps until the upper bound of the number of basis functions, $r_{\text{max}}$, is reached, or the ELMSE does not change substantially. In practice, we recommend using both together as a stopping criterion, and $r_{\text{max}}$ can be chosen as large as computational constraints allow. When computational limits are less constrained, users can set $r_{\text{max}}$ larger, or even let the stopping criterion depend solely on the change in ELMSE.

\begin{algorithm}
	\caption{Forward basis function selection.}	\label{algorithm:basis}
	\begin{algorithmic}[1]
	
	 \State {\bf Input:} Observed data $\{Y(\bs): \bs\in\mathcal{S}_O\}$ with $\mathcal{S}_O$ denoting the set of observation locations, and candidate centers $\mathcal{S}^*\equiv\{ \mathbf{s}_i \in \mathcal{D}: i=1, \ldots, r_*\}$ with fixed and finite $r_*$. 
	 
	 \noindent // \textit{Notice that the location $\bs$ is a generic notation to denote a location where a data value is obtained. The data are not necessarily defined at BAU levels. When data are at coarse spatial resolution, the residuals and related calculations are obtained at the same resolution correspondingly.}
	 \State {\bf Initialization:} $i\gets 1$; a starting set of $r^{(i)}$ basis functions with centers $\mathcal{C}^{(i)}\equiv \{\bc_{1,1},\ldots, \bc_{1,r^{(i)}}\}$ and corresponding bandwidths $\mathcal{B}^{(i)}\equiv \{b_{1,1},\ldots, b_{1,r^{(i)}}\}$.
	 \While{ the stopping criterion is not satisfied} 
	 \State Fit the FRK model with the current basis functions.
	 \State Calculate the pseudo-residuals: $D^{(i)}(\bs)= Y(\bs)-\hat\mu(\bs) - \bS^{(i)}(\bs)\hat\boeta$, for all $\bs\in \mathcal{S}_O$.
	 \ForAll{$\bs\in \mathcal{S}^*$}
	 \State Perform a local semivariogram analysis to obtain the effective range, $d(\bs)$.
	 \State Define the neighborhood $\mathcal{N}(\bs)\equiv\{\bu: \bu\in\mathcal{S}_O\text{, and }\|\bs-\bu\|\leq d(\bs)\}$ and calculate the empirical local mean squared error (ELMSE): $L^{(i)}(\bs)\equiv \var\{{D}^{(i)}(\bu): \bu\in \mathcal{N}(\bs)\} + (\text{ave}\{{D}^{(i)}(\bu): \bu\in \mathcal{N}(\bs)\})^2$.
	 \EndFor
	 \State Calculate $L_0^{(i)}$, a cutoff based on $\{L^{(i)}(\bs): \bs\in \mathcal{S}^*\}$. For example, the 90th percentile of $\{L^{(i)}(\bs): \bs\in \mathcal{S}^*\}$. 
	 \State Define the set of \textit{potential} centers in the $i$-th iteration: $\mathcal{PC}^{(i)}=\{\bs: L^{(i)}(\bs)\geq L_0^{(i)},\,\, \bs\in \mathcal{S}^*\}$, and define the set of \textit{new} centers $\mathcal{NC}^{(i)}\gets\emptyset$, the empty set. Correspondingly, the set of \textit{new} bandwidths $\mathcal{NB}^{(i)}\gets \emptyset$.
	 \ForAll{$\bs\in \mathcal{PC}^{(i)}$}
	 \If {$\|\bs-\bu\|\geq \gamma(\bu)$ for \textit{all} $\bu\in \mathcal{NC}^{(i)}$, where $\gamma(\bu)$ denotes the corresponding separation distance, here chosen to be two-thirds of the effective range, $\gamma(\bu)\equiv 2d(\bu)/3$.}
	 \State Add $\bs$ into $\mathcal{NC}^{(i)}$, with its corresponding bandwidth, the effective range $d(\bs)$, added into $\mathcal{NB}^{(i)}$.
	 \EndIf
	 \EndFor
	 \State Update: $\mathcal{C}^{(i+1)}\gets \mathcal{C}^{(i)}\cup \mathcal{NC}^{(i)}$, and $\mathcal{B}^{(i+1)}\gets \mathcal{B}^{(i)}\cup \mathcal{NB}^{(i)}$.
	 \State $i\gets i+1$
	 \EndWhile
	\end{algorithmic}
\end{algorithm}

The step-by-step procedure is described in Algorithm 2. In local variogram analysis, a parametric variogram function, such as the exponential function, is fitted using data in a small neighborhood as in \cite{Hammerling2012} via weighted least squares or maximum likelihood estimation \citep{Cressie1985}.  \cite{Tadic2015} give a less user-specified way to define the neighborhood, but their method is computationally more complicated due to sampling based on pairwise distances. We nominally set the separation distance to be two-thirds of the effective range, such that the shortest distance between centers of two basis functions added within the same iteration will be no less than the 1.5 times either of their bandwidths. This is motivated by the suggestion in \cite{Cressie2008} regarding overlap between supports of basis functions. To specify the candidate centers $\mathcal{S}^*$, we choose a set of grid points that covers the spatial domain, or simply set them to be the observation locations, $\mathcal{S}_O$. In our simulation study presented in Section~\ref{sec: toy example for basis selection algorithm}, we set $\mathcal{S}^*=\mathcal{S}_O$, and the empirical results show that this choice is robust against large gaps in observations, and gives improved predictive performance. In practice, to maintain computational efficiency, only a small number, typically no more than a few hundred, of basis functions are used \citep{Cressie2008}. The $\boldsymbol \xi$ term in the CAR component of FGP is designed to capture the remaining variation. If the set of basis functions were able to capture the spatial variation completely,  the CAR model in $\boldsymbol \xi$ would reduce to the special case with the spatial dependence parameter $\gamma=0$, or approximately zero. By placing basis functions at locations where the current model fit is poor, and choosing bandwidths adaptively based on information in their local semivariograms, we expect to iteratively adapt to nonstationary spatial variability. Compared to methods in \cite{Katzfuss2013} and \cite{Konomi2014} that require computationally intensive procedures including reversible jump Markov Chain Monte Carlo, our method is simple, intuitive and well-suited to parallel computing environments, since local variogram fitting can be done in parallel.

In summary, the following inputs are required to implement Algorithm 2: the initial set of basis functions, the set of candidate centers $\mathcal{S}^\ast$, the cutoff value for the ELMSEs, and the stopping criterion. The initial set of basis functions can be chosen to be a small number of equally-spaced basis functions whose centers are from a regular coarse grid over the spatial domain. A default choice of $\mathcal{S}^\ast$ is the set of observation locations. In our numerical studies, we choose the 90th percentiles of the ELMSEs as the cutoff value. This cutoff will affect how many basis functions will be added at each iteration of the forward basis function selection algorithm, but the predictive accuracy is not sensitive to the choice of this cutoff. For the stopping criterion, we recommend stopping the algorithm either when the upper bound on the number of basis functions, $r_{\text{max}}$, is reached or when the ELMSEs do not change significantly. In the numerical examples, we set the threshold to be 0.01. Moreover, Algorithm 2 requires the user to choose the type of basis function. In all numerical studies, we use one particular form of the Wendland basis function, but others such as bisquare can also be used. The impact of using different types of basis functions can be assessed via model selection criteria including BIC or cross validation. However, a thorough empirical and theoretical comparison is beyond the scope of this work.


\section{Simulation Studies}\label{sec: simulation examples}
We present two simulation studies in this section. Section~\ref{subsec:downscaling} presents an  illustration of the downscaling framework, and the importance of handling the change-of-support problem. In Section~\ref{sec: toy example for basis selection algorithm}, we demonstrate how our algorithm for forward basis function selection works, and its superior performance compared to widely equally-spaced basis functions.

\subsection{A Synthetic Example for Statistical Downscaling} \label{subsec:downscaling}
\begin{figure}[htbp]
\begin{center}
\makebox[\textwidth][c]{ \includegraphics[width=1.0\textwidth, height=0.35\textheight]{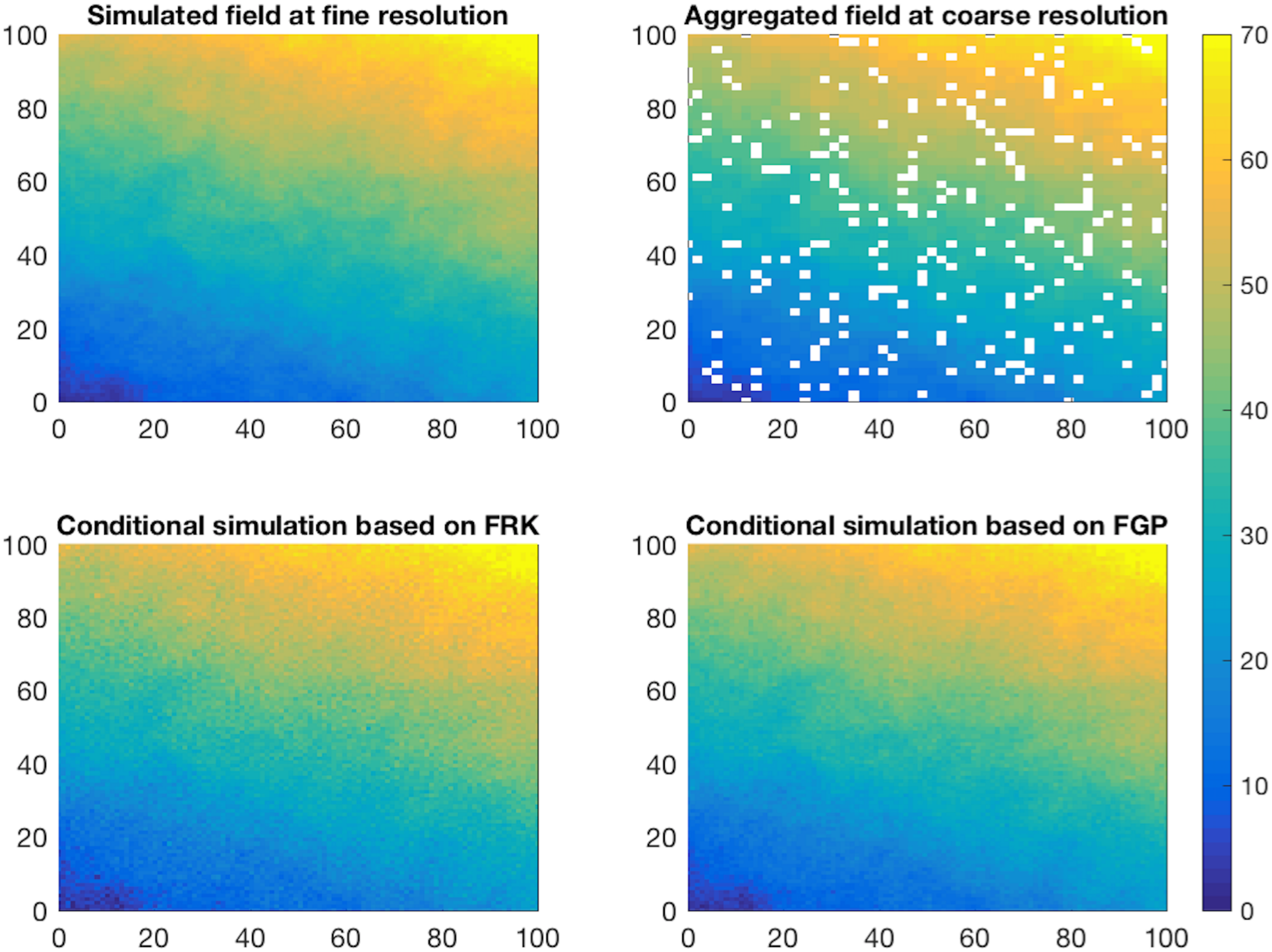}}
\caption{Simulated data and downscaling results from FRK and FGP over the entire domain $[0, 100] \times [0, 100]$. The top-left panel shows $\bY$, the simulated data at fine-resolution. The top-right panel plots the ``synthetic observations'' at coarse resolution, after randomly taking out 10\% of $\widetilde{\bY}$ in the white regions. Bottom panels show the downscaled fields from FRK (bottom-left) and FGP (bottom-right). \vspace{-0.2cm}}
\label{fig: all_sim_expmodel}
\end{center}
\end{figure}

In this section, we use synthetic data to illustrate our the statistical downscaling method. First, we simulate a fine-resolution dataset, and aggregate it to form a coarse-resolution dataset. Then, we downscale this coarse-resolution dataset to obtain a fine-resolution field, which we compare to the originally simulated fine-resolution data. Specifically, we simulate from a Gaussian process with mean $\mu(\mathbf{s})=\bX(\mathbf{s})^T \bbeta$ and  exponential covariance function $c(h)=\sigma^2\exp(-h/\rho) + \sigma^2_{\epsilon} I(h=0)$ at $N=100\times 100$ regular BAUs in a $[0, 100] \times [0, 100]$ domain with $\mathbf{s}\equiv (x, y)^T$. Here, $\bX(\mathbf{s})=(1, x, y)^T$ with $\bbeta=(2, 0.5, 0.2)^T$, $\sigma^2=2$, $\rho=5$, and  $\sigma^2_{\epsilon}=0.2$. The simulated dataset is denoted by $\mathbf{Y}=(Y(\mathbf{s}_1), \ldots, Y(\mathbf{s}_N))^T$ at the $N$ BAUs. Then, the fine-resolution dataset $\mathbf{Y}$ is upscaled to the coarse-resolution dataset at $M=50\times 50$ regular grid cells in the same domain, using the change-of-support property in Eq.~\eqref{eqn: approximate cos}. The resulting coarse-resolution data are referred to as $\widetilde{\mathbf{Y}}$. To assess the quality of the downscaling approach, 10\% of coarse-resolution data are randomly held out. These are shown in white regions in Figure~\ref{fig: all_sim_expmodel}. The remaining 90\% of coarse-resolution data are treated as ``synthetic observations''. 

Based on synthetic observations, we implement four methods:
\begin{itemize}[noitemsep,topsep=4pt]
\item[(1)] kriging using the true parameters in an exponential covariance function model that accounts for change of support;
\item[(2)] kriging based on estimated parameters in the exponential covariance function model \textit{without} handling change-of-support problem. That is, the coarse-resolution data are treated as point-level ones at the centers of coarse-resolution grid cells;
\item[(3)] kriging based on FRK presented in Section~\ref{sec: spatial model}; 
\item[(4)] kriging based on FGP presented in Section~\ref{sec: spatial model}. 
\end{itemize}
These methods are referred to as EK, PK, FRK, FGP, respectively. Note that the kriging predictor is indeed the conditional mean of the distribution of the true field given the data. The parameters in FRK and FGP are estimated using maximum likelihood methods given in Section~\ref{subsec:para_est}. The basis functions are chosen at three different resolutions with $152=4^2+6^2+10^2$ equally-spaced centers. The proximity matrix in FGP is chosen based on first-order neighborhood structure. 

\begin{figure}[htbp]
\begin{center}
\makebox[\textwidth][c]{ \includegraphics[width=1.0\textwidth, height=0.35\textheight]{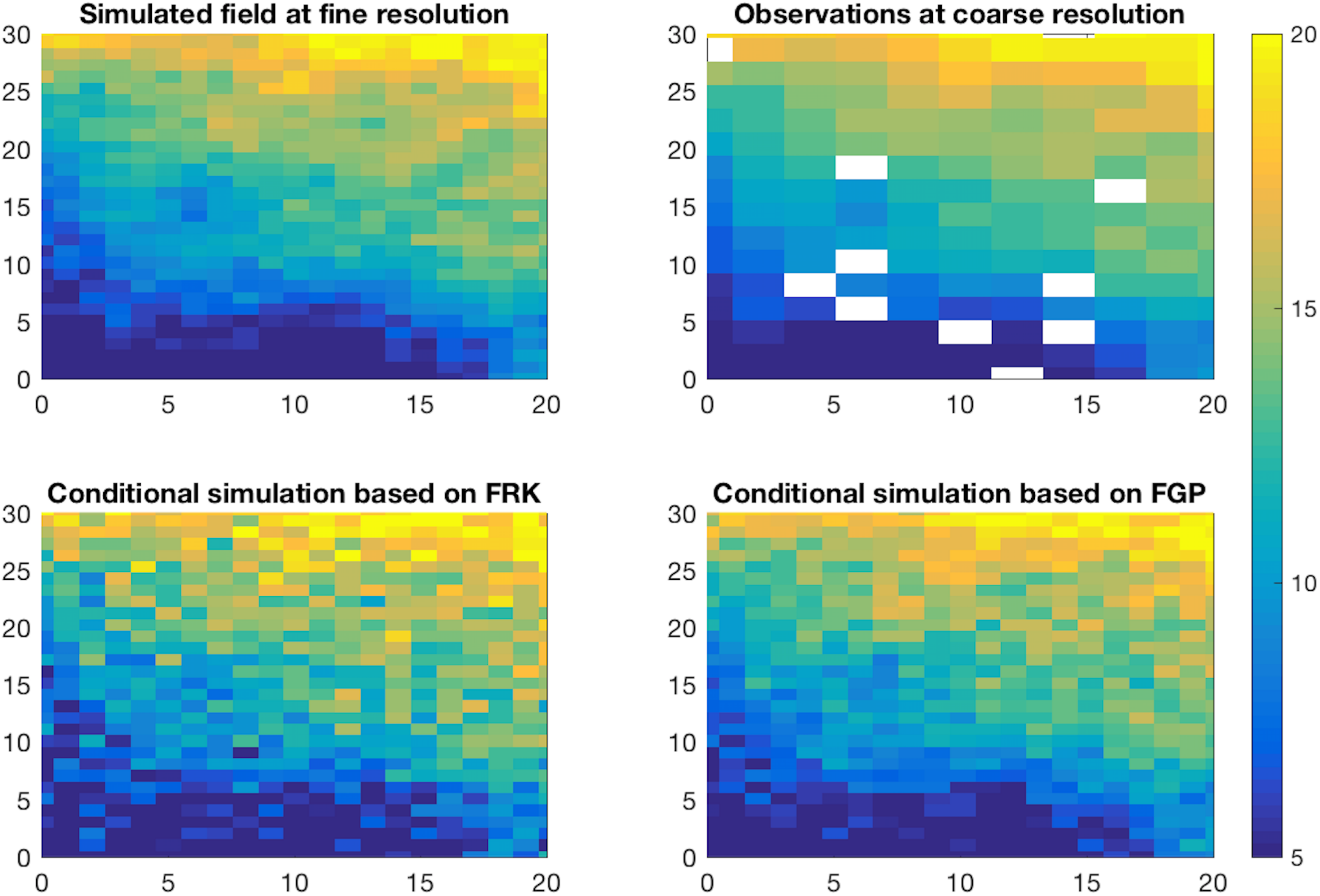}}
\caption{Simulated data and downscaling results from FRK and FGP, with a zoomed-in view over the subregion $[0, 20]\times [0, 30]$. The top-left panel shows $\bY$, the simulated data at fine-resolution. The top-right panel shows the ``synthetic observations'' at coarse resolution. Bottom panels show the downscaled fields from FRK (bottom-left) and FGP (bottom-right).}
\label{fig: zoomin_sim_expmodel}
\end{center}
\end{figure}

Spatial predictions are made at all fine-resolution BAUs, and we compare the predictions from the four methods with the ``truth'', $\bY$, by calculating the mean-squared-prediction errors (MSPEs). We see from Table~\ref{table: prediction_sim_results} that EK gives the best results as expected, since it uses the true parameters and covariance function model, and handles change-of-support problem. PK does not handle change-of-support problem, and treats coarse-resolution data incorrectly as being at fine resolution. The predictions from PK are much worse than those from FRK and FGP which do account for change of support. Between FRK and FGP, the latter gives better predictive performance, as expected, since its model is more flexible. Figure~\ref{fig: all_sim_expmodel} visualizes the fine-resolution true fields and synthetic observations at coarse-resolution, together with the downscaled fields from FRK and FGP, over the entire region. Figure~\ref{fig: zoomin_sim_expmodel} shows the results zoomed-in on a  $[0, 20]\times [0, 30]$ region. The downscaled fields from both FRK and FGP mimic the pattern presented in the true fine-resolution field, but the FGP field is closer to $\bY$.

\begin{table}[htbp]
\centering
\normalsize
   \caption{Summary of results for spatial predictions based on coarse-resolution data using four methods: EK, PK, FRK, and FGP, respectively. ``COS'' stands for change of support, which indicates whether the method deals with change of support or not.} 
  {\resizebox{1.0\textwidth}{!}{%
  \setlength{\tabcolsep}{3.0em}
   \begin{tabular}{l r r r r} 
   \toprule 
 Method	 & EK & PK & FRK & FGP \\ \noalign{\vskip 2.5pt}  \hline \noalign{\vskip 2.5pt}
COS & Yes & No & Yes & Yes \\ \noalign{\vskip 1.5pt}
MSPE & 0.457 & 0.623 & 0.570 & 0.477 \\ 
 \bottomrule
   \end{tabular} 
   }}
   \label{table: prediction_sim_results}
\end{table} 
\vfil
\vfil
\vfil

\subsection{A Toy Example for the Forward Basis Function Selection Algorithm} \label{sec: toy example for basis selection algorithm}
\begin{figure}[htbp]
\begin{center}
\makebox[\textwidth][c]{ \includegraphics[width=1.0\textwidth, height=0.4\textheight]{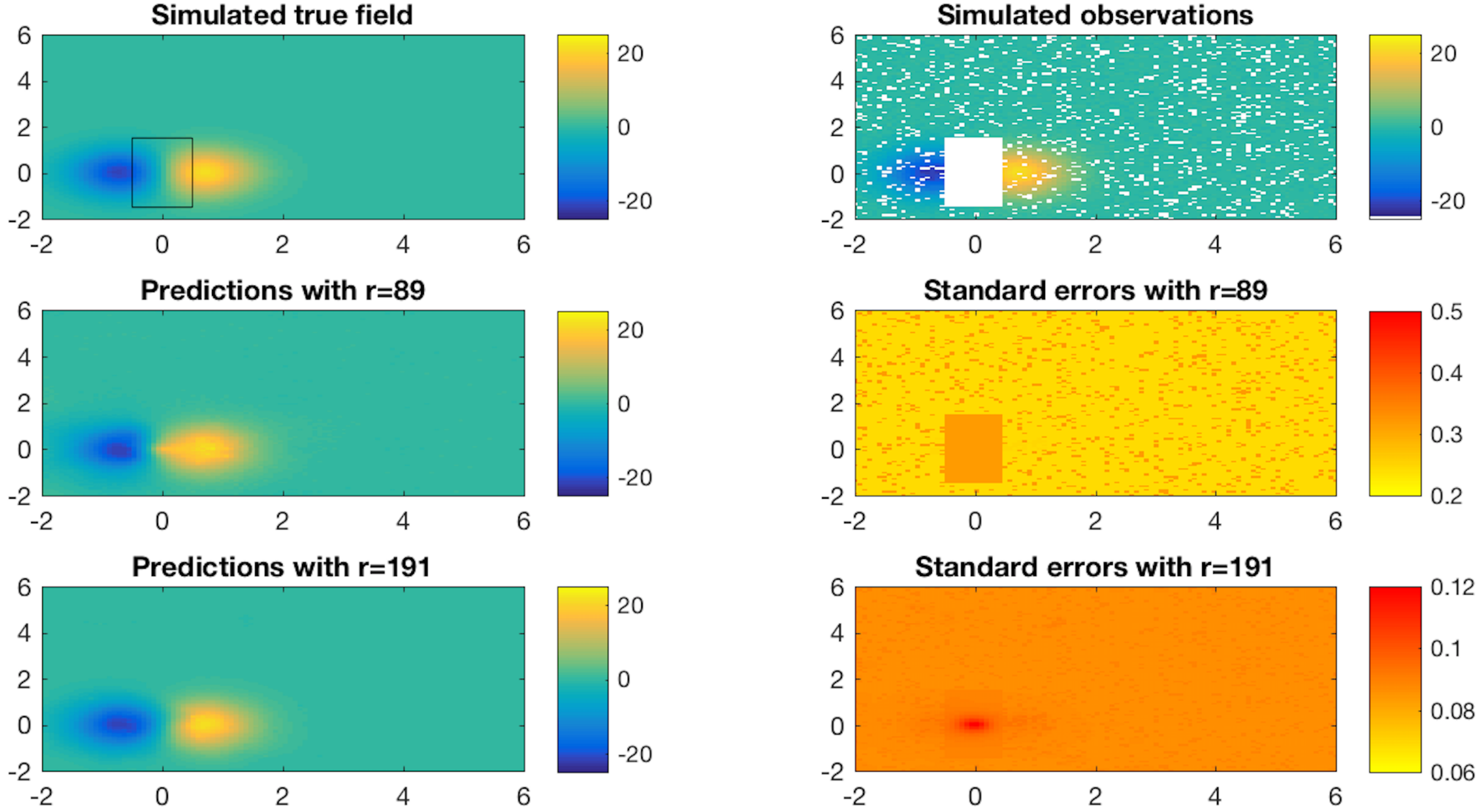}}
\caption{Predictions with adaptive basis functions from the forward selection algorithm. Top-left panel shows the true deterministic field on a $100 \times 100$ regular grid; top-right panel shows the observations after removing the prediction locations, $\mathcal{S}_1 \cup \mathcal{S}_2$. The white box in the top-right panel shows the missing-by-design locations, $\mathcal{S}_1$, and other white locations show the missing-at-random locations, $\mathcal{S}_2$. The middle and bottom panels are predictions for the underlying true field, and associated standard errors with $r=89$ and 191 basis functions, respectively.\vspace{-0.5cm}}
\label{fig: pred_AD_sim}
\end{center}
\end{figure}

Here we give an example to illustrate the method and performance of our algorithm for forward basis function selection. Consider the deterministic function $f(\mathbf{s})=50x\exp(-x^2-y^2)$ for $\bs\equiv (x, y)^T$ in the domain $\mathcal{D}\equiv [-2, 6] \times [-2, 6]$. This function has two localized features inside the subregion $[-2, 2]\times [-2, 2]$, and is almost zero everywhere else; see the top-left panel in Figure~\ref{fig: pred_AD_sim}. To create the synthetic true field, we generate function values of $f(\cdot)$ on the $100\times 100$ regular grid covering the domain. We add a Gaussian white noise term with mean zero and variance $\sigma^2_{\epsilon} = 0.01 \hat{\sigma}^2_f$ at each grid cell, where $\hat{\sigma}^2_f$ is the empirical variance of the function $f(\cdot)$ evaluated at these $100\times 100$ regular grid cells. We hold out data on a small block region $\mathcal{S}_1\equiv [-0.5, 0.5] \times [-1.5, 1.5]$, referred to as ``missing-by-design'' locations. We also hold out data at randomly selected $10\%$ of the remaining  grid cells $\mathcal{S}_2$, referred to as ``missing-at-random'' locations. The remaining 90\% of data are treated as ``observations" for which their locations are denoted by $\mathcal{D}_o \equiv \mathcal{D} \setminus (\mathcal{S}_1 \cup \mathcal{S}_2)$; see the top-right panel in Figure~\ref{fig: pred_AD_sim}.

\begin{figure}[htbp]
\begin{center}
\makebox[\textwidth][c]{ \includegraphics[width=1.0\textwidth, height=0.4\textheight]{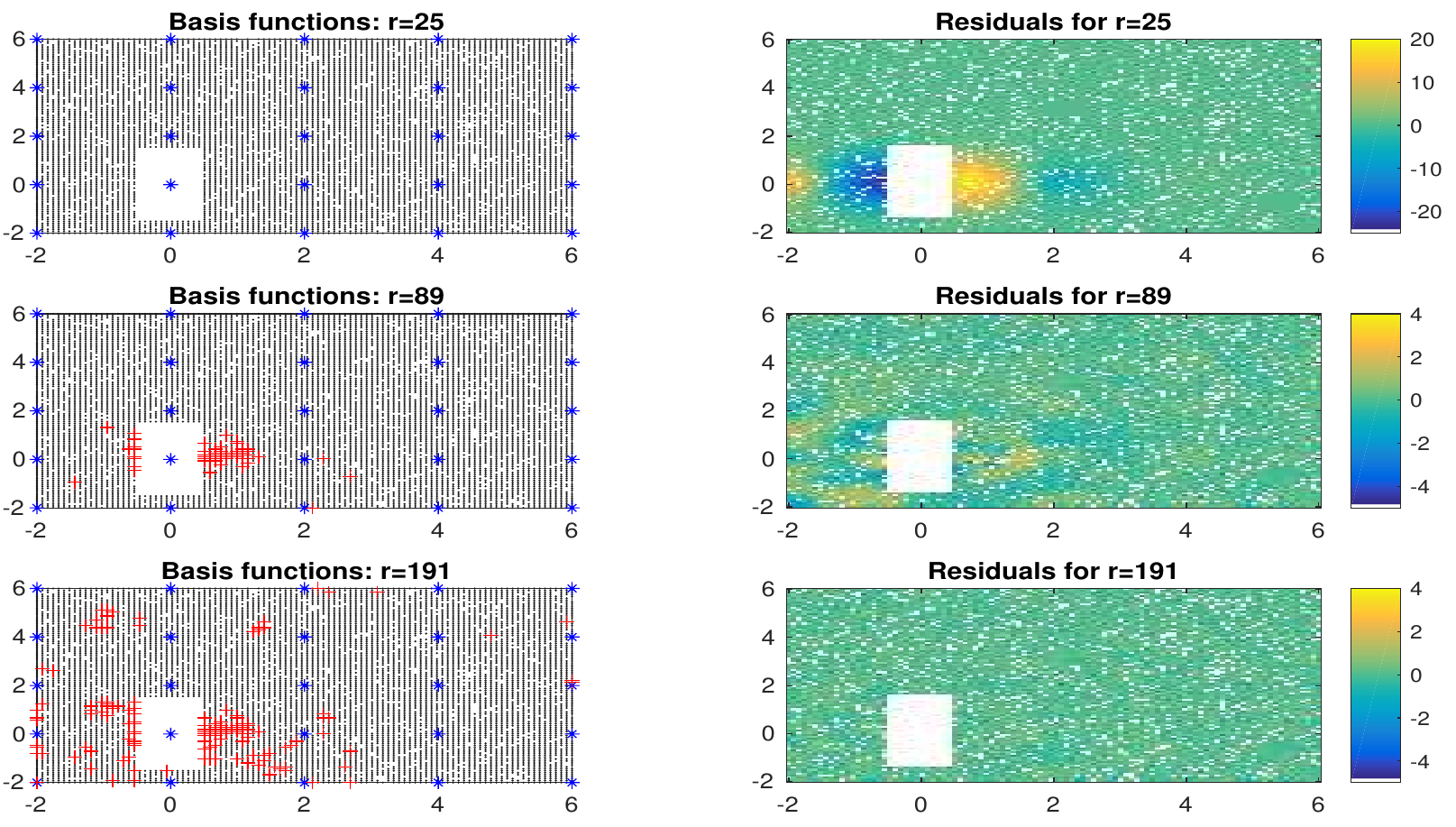}}
\caption{Adaptive basis functions and corresponding residuals for simulation example. The asterisks in three left panels represent 25 initial basis centers, and the dots show the observation locations. The plus signs represent new basis centers added using the forward selection algorithm. The three right panels show the residuals in the forward selection algorithm with different numbers of basis functions $r=25, 89, \text{and } 191$, respectively.\vspace{-1cm}}
\label{fig: AD_basis_sim}
\end{center}
\end{figure}

\begin{figure}[htbp]
\begin{center}
\makebox[\textwidth][c]{ \includegraphics[width=1.0\textwidth, height=0.25\textheight]{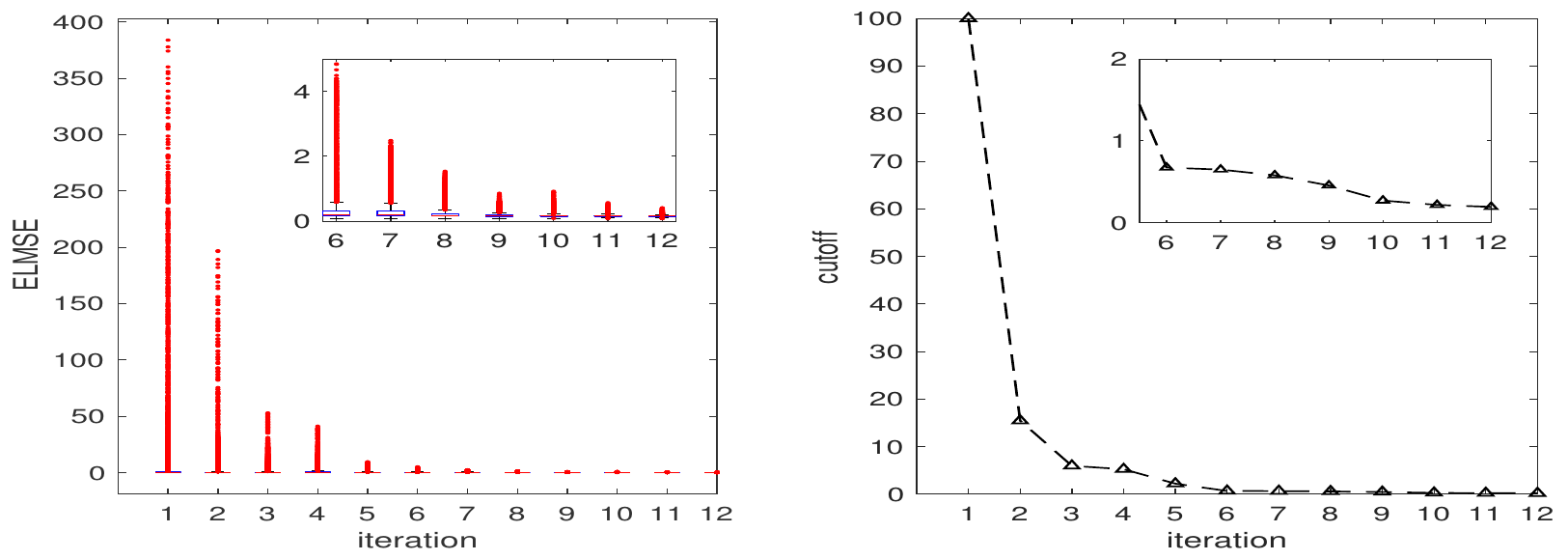}}
\caption{Diagnostics of the forward selection algorithm at each iteration for simulation example. The left panel shows the box plot of ELMSEs at each iteration of the forward selection algorithm. In the right panel, the triangles $\bigtriangleup$ show the 90th percentile of the ELMSEs, i.e., cutoff, at each iteration of the forward selection algorithm. \vspace{-1cm}}
\label{fig: measures_sim}
\end{center}
\end{figure}

The initial set of basis functions is chosen to be a set of 25 Wendland basis functions with centers equally spaced over the domain, as shown in the top-left panel of Figure~\ref{fig: AD_basis_sim}. The corresponding bandwidths are 1.5 times the shortest distance among these 25 centers, as suggested in \cite{Cressie2008}. In the forward selection algorithm, we set the maximum number of basis functions to be $r_{\text{max}}=200$, and set the stopping criterion to be the time that the absolute difference between the cutoff values $L^{(i)}_0$ at two consecutive iterations is at or below $0.01$. In our example, the forward selection algorithm stops after 12 iterations, resulting in a total of 191 basis functions (including the original 25). The middle-left and bottom-left panels of Figure~\ref{fig: AD_basis_sim} plot the centers of basis functions added at the eighth and 12th iterations, respectively. The corresponding pseudo-residuals for these two iterations are shown in the middle-right and bottom-right panels, respectively. We see that basis functions are placed where the variabilities of pseudo-residuals are large. After adding the new basis functions, the model fits the data more closely. Figure~\ref{fig: pred_AD_sim} shows spatial predictions and associated standard errors after the eighth and 12th iterations of the procedure. It is obvious that the predicted field is closer to the true field using 191 basis functions (i.e., after all 12 iterations), compared to those from only the eighth iteration. This demonstrates that our algorithm adds basis functions in a way that improves predictive performance. Figure~\ref{fig: measures_sim} shows how the associated ELMSEs $\{L^{(i)}(\bs)\}$ and the cutoffs, $L^{(i)}_0$, change with iteration for $i=1,\ldots, 12$. Observe that both ELMSE and $L^{(i)}_0$ decrease as basis functions are added. 

To further demonstrate the advantage of our adaptive basis function approach, we compare our results with the more commonly used, equally-spaced basis function approach. \cite{Cressie2008}, \cite{Nguyen2012}, and \cite{Nguyen2014} suggest using compactly-supported basis functions at several resolutions, but choose basis functions at the same resolution to be equally-spaced. When we fit a model with equally-spaced basis functions, we call the method ``simple'', and we call the method  ``adaptive'' if we instead use the adaptive basis functions from our forward algorithm. Here, we use equally-spaced basis functions from three resolutions, giving a total of $151(=5^2+7^2+9^2-4)$ basis functions, with four basis functions removed where data are sparse \citep{Zhu2015, Zammit2017}. We also add additional equally-spaced basis functions at the next-finer resolution, resulting in a total of $264(=5^2+7^2+9^2+11^2-12)$ basis functions. 

We compare the predictive performance of the simple and adaptive methods by looking at their corresponding mean square prediction errors (MSPEs). As shown in Table~\ref{table: MSPE for basis comparison}, simple FRK with more basis functions ($r = 264$) gives better predictions at both missing-by-design and missing-at-random locations, compared to simple FRK with only $r=151$ basis functions. However, using only $r=191$ adaptive basis functions, adaptive FRK gives much better predictions than simple FRK with more basis functions. Specifically, the MSPE over all missing locations from adaptive FRK with 191 basis functions is only about 36\% of that from simple FRK with 264 basis functions. 

We then fit FGP in which a CAR model is assumed for the random vector, $\bxi$. The proximity matrix for FGP is specified by assuming a parsimonious first-order neighborhood structure. Recall that FGP reduces to FRK when the spatial dependence structure parameter is equal to zero, and thus FGP is more flexible \citep{Ma2017}. With 191 adaptive basis functions,  adaptive FGP gives the best predictive performance overall. Compared to simple FRK with $r=264$ functions, its MSPE, over the missing-by-design locations, is less than one-third that of simple FRK. We see even more improvement at the missing-at-random locations: the MSPE of adaptive FGP over these locations is 0.015, compared to 0.477 from simple FRK with 264 basis functions.  

For comparison, we perform local kriging \citep{Haas1990, kitanidis_1997}. We choose a neighborhood composed of 6-by-6 pixels for each location in $\mathcal{S}_2$, and a neighborhood of 15-by-15 pixels for each grid cell for locations in $\mathcal{S}_1$, and perform kriging with an exponential covariance function. From Table~\ref{table: MSPE for basis comparison}, we see that over $\mathcal{S}_1$ (i.e, data missing in a contiguous region), both adaptive FRK and adaptive FGP outperform local kriging substantially, while they give comparable results over $\mathcal{S}_2$ (i.e., data missing at random). Note that local kriging is based on a moving window of nearest observations. When prediction is made over a contiguous missing region, the majority of prediction locations in that region will depend mostly on the same set of nearby observations. In contrast, adaptive FRK and adaptive FGP, with appropriately chosen basis functions, are better able to capture spatial dependence structures, and give better prediction results. Our finding here is also consistent with that of \cite{Shi2007}: globally valid, flexible models such as FRK and FGP outperform local kriging and other fast non-statistical spatial prediction methods, such as Inverse Distance Weighting (IDW)  and Nearest Neighbors Smoothing (NNS) for data that have contiguous missing values. Moreover, local kriging only gives \textit{marginal} inference at \textit{each} location \textit{separately}, but FRK and FGP, as global models over the entire spatial domain, are able to provide \textit{joint} inference at \textit{all} locations of interest. This is the key to properly quantify uncertainties with a globally valid spatial process model via conditional simulation.

We now report on computational time for methods in this simulation study. The forward basis function selection algorithm took about 70 seconds on a Macbook Pro with a 2.8-GHz Intel Core i7 processor. The parameter estimation and prediction took about 12, 137, 35 seconds for adaptive FRK, adaptive FGP, and  local kriging, respectively. We can see that here FRK and local kriging are faster than FGP, but FGP gives smaller MSPE, overall. 

\begin{table}[htbp]
\centering
\normalsize
   \caption{Numerical results for comparing local kriging, FRK, and FGP.  Here, $r$ denotes the total number of basis functions in the low-rank component in FRK and FGP. The method is called ``simple'' if equally-spaced basis functions are employed, and ``adaptive'' when basis functions are selected via Algorithm~\ref{algorithm:basis}.}
  {\resizebox{1.0\textwidth}{!}{%
  \setlength{\tabcolsep}{1.5em}
   \begin{tabular}{l c c c c c} 
   \toprule 
\multirow{2}{*}{Method} & \multirow{2}{*}{Local Kriging} & \multicolumn{2}{c}{Simple FRK} & Adaptive FRK  & Adaptive FGP  \\ \noalign{\vskip 1pt} \cline{3-4} \noalign{\vskip 0.5pt}
	& &$r=151$ & $r=264$ & $r=191$ & $r=191$	\\ \noalign{\vskip 1.5pt} \hline \noalign{\vskip 2.5pt}
MSPE($\mathcal{S}_1$) &  6.252 & 7.451 & 6.753 & 2.431 &2.042   \\ \noalign{\vskip 1.5pt}
MSPE($\mathcal{S}_2$) & 0.010 & 0.495 & 0.477 & 0.019 &  0.015  \\ \noalign{\vskip 1.5pt}
MSPE($\mathcal{S}_1\cup\mathcal{S}_2$) &1.929 &2.634 & 2.407 &0.761 &0.638    \\ 
 \bottomrule
   \end{tabular}%
   }}
   \label{table: MSPE for basis comparison}
\end{table}  
		

\section{Application to PCTM}\label{sec: application on pctm}

Atmospheric carbon dioxide ($\text{CO}_2$) is one of the most important greenhouse gases. Numerical models are typically used to study geophysical processes in carbon cycle, and to assess future climate change \citep[e.g.,][]{Friedlingstein2006}. Specifically, high-resolution atmospheric $\text{CO}_2$ fields over the globe produced by numerical models are often used to study atmospheric chemistry and dynamics. Global numerical models typically generate atmospheric $\text{CO}_2$ at relatively coarse-resolution, say, $100\sim500$ $\text{km}$  grid cells. On the other hand, since surface-level emissions are more relevant to urban environments where the majority of people reside, high-resolution surface observing networks are required to monitor greenhouse gas emissions, and to devise mitigation strategies \citep[e.g.,][]{Shusterman2016}. OSSEs can be used to design these observing networks, and to evaluate data assimilation algorithms that combine their data with space-based observations like those from Japan's Greenhouse Gases Observing Satellite (GOSAT) and NASA's Orbiting Carbon Observatory-2 (OCO-2). In this section, we demonstrate how our downscaling framework can be used to construct high-resolution NR fields for OSSEs. 

In our study, we downscale surface $\text{CO}_2$ concentrations generated by the PCTM/GEOS-4/CASA-GFED model. This numerical model has been widely used to study global CO$_2$ and to evaluate mapping algorithms \citep[e.g.,][]{Parazoo2011, Hammerling2012, Zhang2014}. We use PCTM to simulate global atmospheric CO$_2$ concentrations, with unit parts per million, at the surface on January 3, 2006. The spatial resolution of this output is $1^{\circ}$ latitude by $1.25^{\circ}$ longitude, which results in $M=181\times 288=52,128$ grid cells over the globe. Statistical downscaling is performed on this PCTM output to construct a global high-resolution surface $\text{CO}_2$ field on equal-area hexagonal grid cells, with 30 km intercell distances. These hexagonal grid cells are obtained from the Discrete Global Grid software \citep[DGG,][]{Sahr2003,Stough2014}, and the hexagons are used as the BAUs in our study. The surface of the Earth is uniformly tiled by $N=655,362$ BAUs. Exploratory analysis suggests a linear trend based on latitude, and nonstationary spatial dependence structure. In what follows, we describe how adaptive basis functions are selected for PCTM output with our forward selection algorithm, and present the downscaled high-resolution NRs based on both FRK and FGP.

\begin{figure}[htbp]
\begin{center}
\makebox[\textwidth][c]{ \includegraphics[width=1.0\textwidth, height=0.5\textheight]{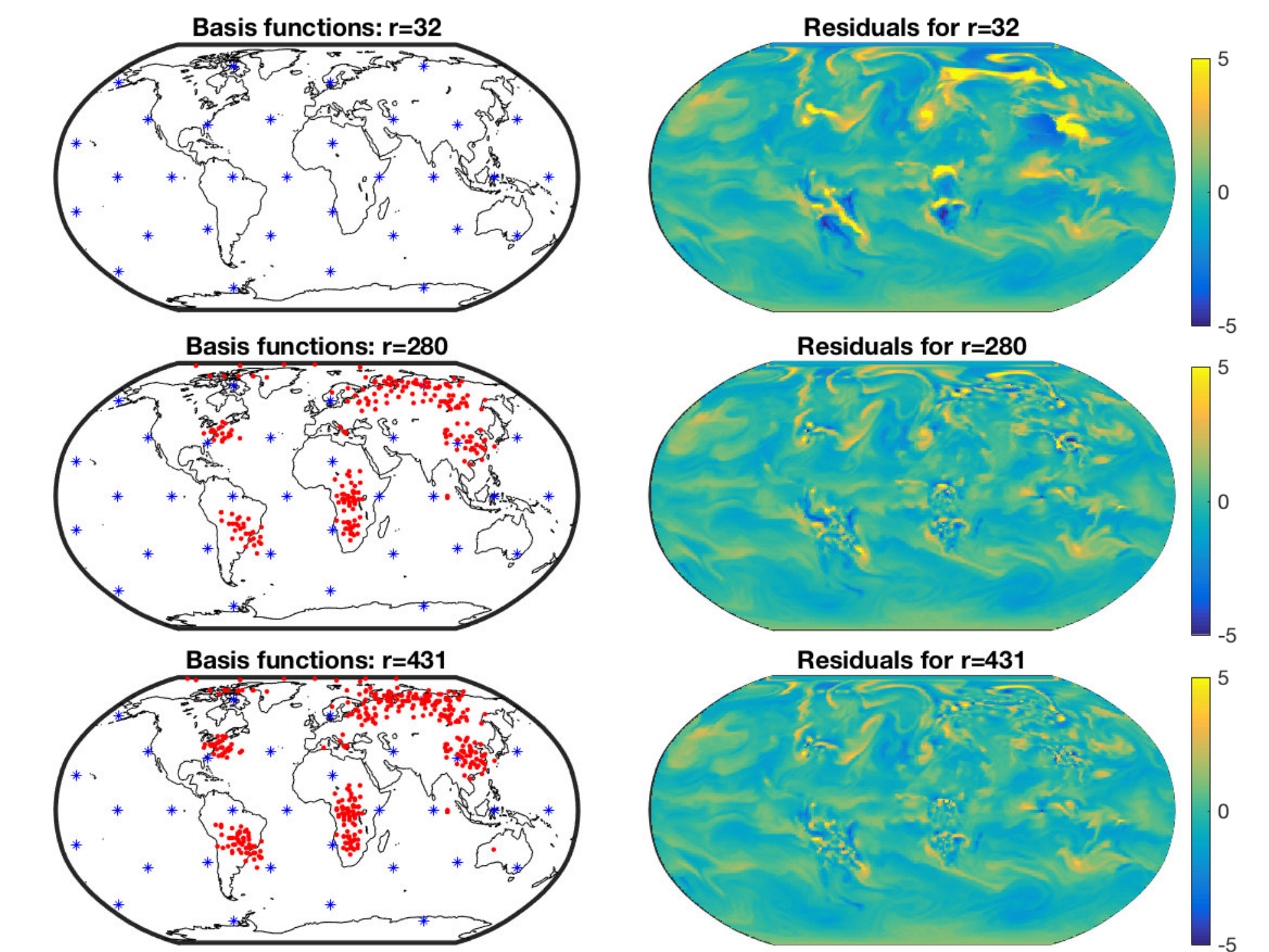}}
\caption{Adaptive basis functions and corresponding residuals for PCTM output. The asterisks $\ast$ in three left panels represent 32 initial basis centers. The dots $\bullet$ represent new basis centers added using the forward selection algorithm. The three right panels show the residuals in the forward selection algorithm with different number of basis functions $r=32, 280, \text{and } 431$, respectively. \vspace{-0.1cm}}
\label{fig: AD_basis_PCTM}
\end{center}
\end{figure}

\begin{figure}[htbp]
\begin{center}
\makebox[\textwidth][c]{ \includegraphics[width=1.0\textwidth, height=0.25\textheight]{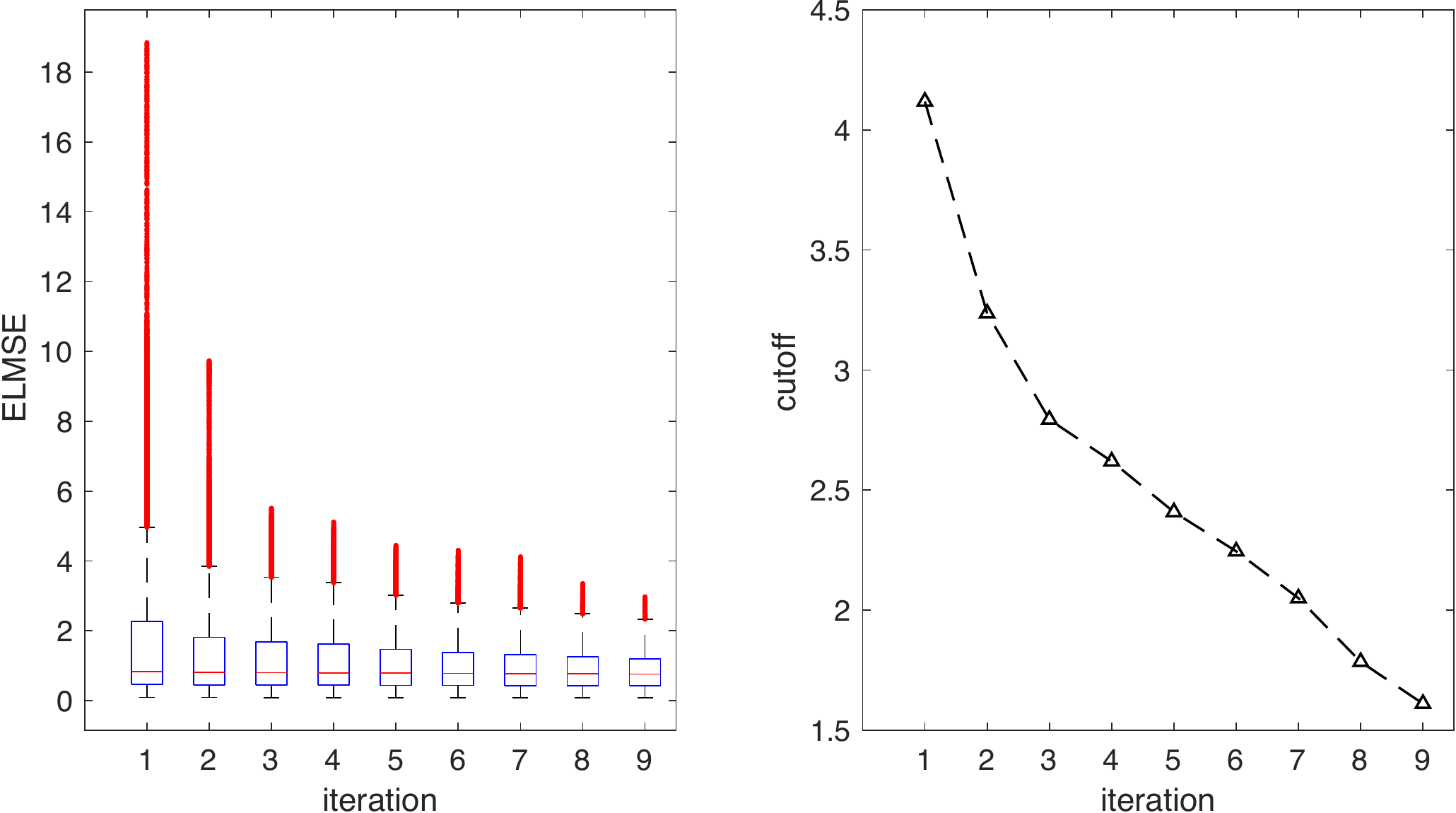}}
\caption{Diagnostics of the forward selection algorithm at each iteration for PCTM outputs. The left panel shows the boxplot of ELMSEs at each iteration of the forward selection algorithm. In the right panel, the triangles $\bigtriangleup$ show the 90th percentile of ELSMEs, i.e., cutoff, at each iteration of the forward selection algorithm.\vspace{-0.5cm}}
\label{fig: measures_PCTM}
\end{center}
\end{figure}

To apply the forward basis selection algorithm, we begin with a set of $32$ equally-spaced basis functions with radii 6241.1 km obtained from \cite{Cressie2008}. The centers of these 36 basis functions are shown as blue asterisks in the left panels of Figure~\ref{fig: AD_basis_PCTM}. We set the collection of candidate centers to the set of the centroids of all grid cells. In this application, pseudo-residuals are at the same coarse resolution as PCTM output, and local semivariogram analyse are carried out. We use a stopping criterion that requires the number of basis functions not exceed 450, and simultaneously, that  the absolute change in the cutoff value, $L^{(i)}_0$, between two consecutive iterations not exceed 0.01. The forward selection algorithm stops after the ninth iteration, resulting in a total of $r=431$ basis functions. The cutoff at the ninth iteration is $L^{(9)}_0=1.61$, and the difference between the cutoffs at the eighth and ninth iterations is 0.17. Figure~\ref{fig: measures_PCTM} plots the ELMSEs and the cutoffs, $L^{(i)}_0$, as a function of iteration, for $i=1, \ldots, 9$. The cutoff decreases as the number of iterations increases. The algorithm stops after the ninth iteration because of the upper limit on the total number of basis functions. Recall that inference, including parameter estimation and downscaling via conditional simulation, requires inverting $r\times r $ matrices and storing $N\times r$ matrices, thus large $r$ is not desirable.

\begin{figure}[htbp]
\begin{center}
\makebox[\textwidth][c]{ \includegraphics[width=.90\textwidth, height=0.9\textheight]{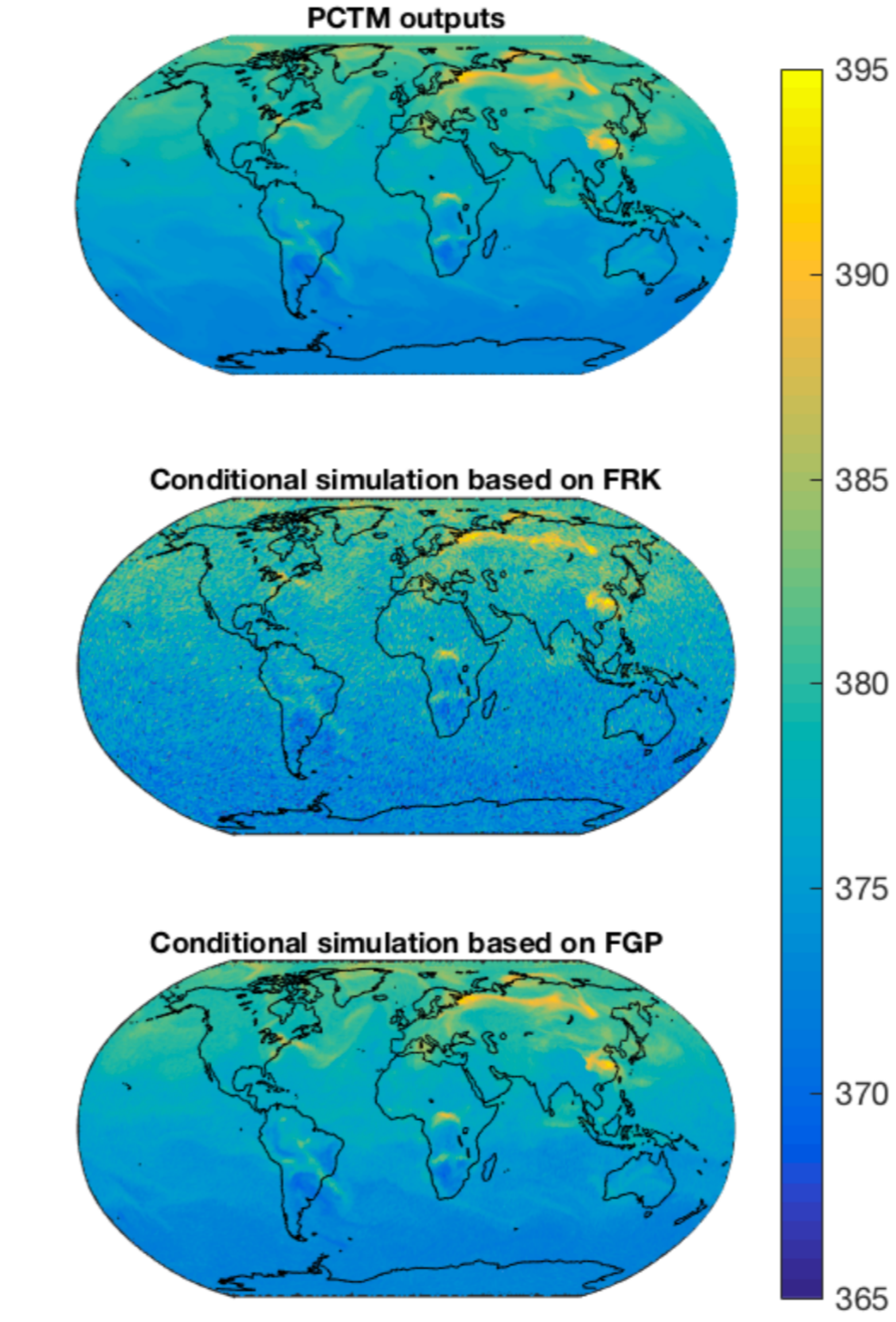}}
\caption{Global map of PCTM outputs and downscaled fields. The top panel shows the surface CO$_2$ concentrations with unit parts per million (ppm) from PCTM at $1^{\circ} \times 1.25^{\circ}$ latitude and longitude; middle and bottom panels show the downscaled fields for surface CO$_2$ based on FRK and FGP at 30 km spatial resolution.}
\label{fig: map of downscaled pctm}
\end{center}
\end{figure}

\begin{figure}[htbp]
\begin{center}
\makebox[\textwidth][c]{ \includegraphics[width=1.0\textwidth, height=0.6\textheight]{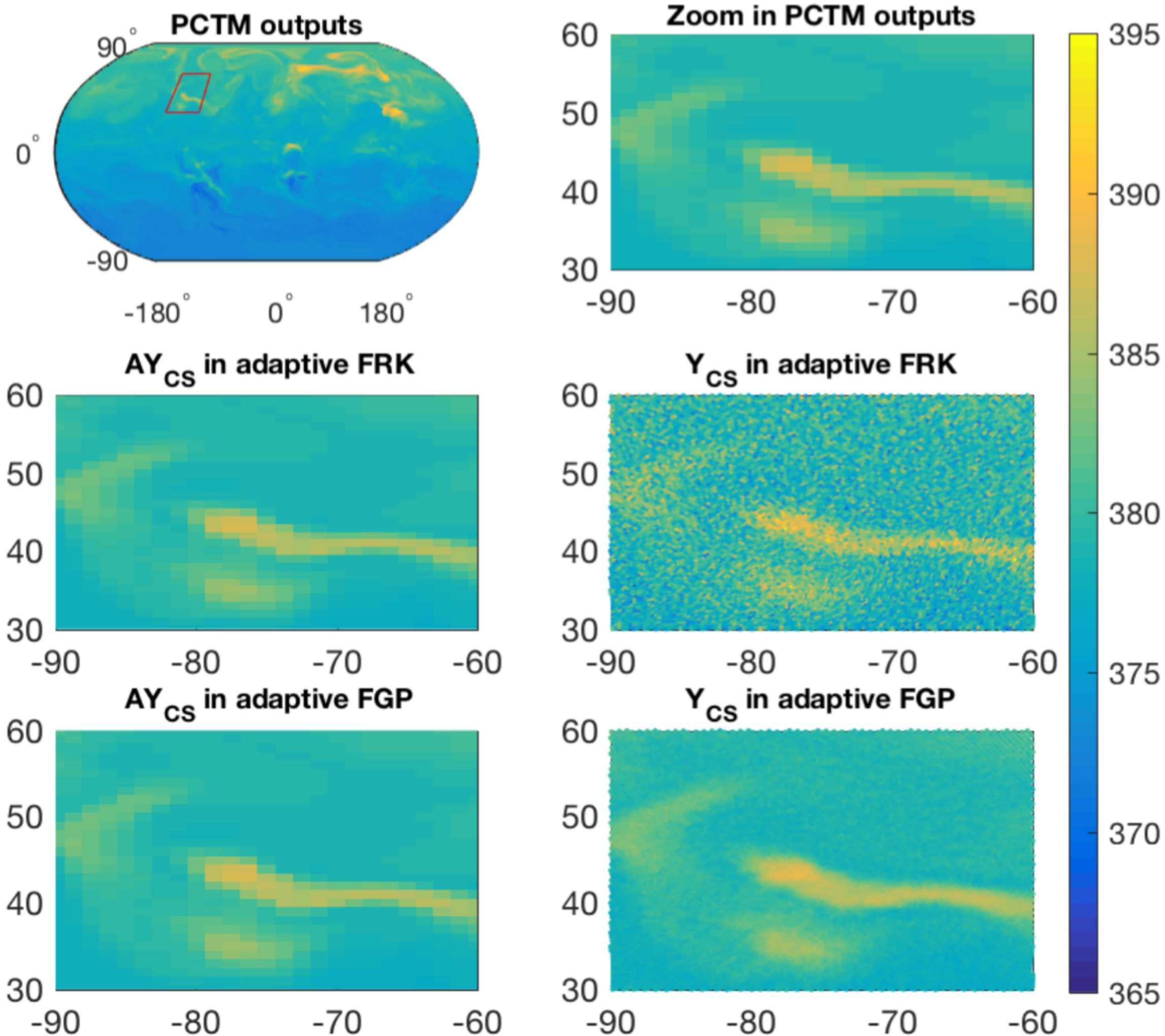}}
\caption{PCTM outputs and aggregated downscaled fields at $1^{\circ} \times 1.25^{\circ}$ resolution, and downscaled fields based on FRK and FGP at 30 km spatial resolution (in ppm). Top-left panel shows the global PCTM outputs with zoom-in region identified by the black rectangle. The top-right panel shows the PCTM outputs for the zoom-in region at 30 km resolution.  Middle-left panel shows the aggregated conditional simulated values based on FRK; middle-right panel shows the conditional simulated values based on FRK. Bottom-left panel shows the aggregated conditional simulated values based on FGP; bottom-right panel shows the conditional simulated values based on FGP.}
\label{fig: zoom-in downscaled pctm}
\end{center}
\end{figure}

With these $r=431$ adaptive basis functions, we implement the downscaling framework based on FRK and FGP. Figure~\ref{fig: map of downscaled pctm} shows the PCTM output over the globe and the downscaled fields: the high-resolution NRs, from conditional simulation based on FRK and FGP, respectively. Although the spatial pattern in PCTM output is maintained by both high-resolution NRs, we judge the NR based on FGP to be superior, since the FRK NR presents a clear ``salt-and-pepper'' artifact, which is not realistic for atmospheric processes. Such ``salt-and-pepper'' artifacts appear more clearly when we zoom into a subregion, as shown in Figure~\ref{fig: zoom-in downscaled pctm}. Specifically, note that when aggregated back to the $1^\circ\times 1.25^\circ$ resolution, the high-resolution NRs from both FRK and FGP match the PCTM output exactly. However, the FGP NR at high resolution does not present the salt-and-pepper artifacts. Using BIC to compare FRK and FGP with 431 adaptive basis functions. We found that BIC$_{FRK}=22.65$ and BIC$_{FGP}=20.37$, which also suggests that FGP performs better than FRK. This is also consistent with the empirical results in our simulation studies that show that adaptive FGP gives better model fit than adaptive FRK. Here, both FRK and FGP are run on an HP Intel Xeon E5-2690 machine with 12 GB memory and four cores at the Ohio Supercomputer Center \citep{OSC}. The adaptive basis function selection algorithm took about 15 minutes. The computation times for parameter estimation are about three minutes for FRK and ten hours in FGP. The latter takes more time because FGP requires additional calculations related to $N\times N$ sparse matrices and numerical optimization to update the spatial dependence parameter in the CAR part of the model within each iteration of the EM algorithm. To generate one downscaled NR, FRK took about one minute, and FGP about four minutes. Therefore, we are able to produce ensembles of high-resolution NRs efficiently for both FRK and FGP. We have tried  implementing local kriging in parallel on the HP Intel Xeon E5-2690 machine with 12 GB memory and 12 cores. Note that local kriging does not define a valid joint predictive distribution for all BAUs. Therefore, the downscaled field from local kriging fails to satisfy the important aggregation requirement. Furthermore, due to the large size of $N=655,362$, it took about 20 hours to run local kriging at all $N$ BAUs. Computation time required for local kriging could be shortened when more computing cores (and thus more extensive parallelization) are available.

\section{Conclusions and Discussion}\label{sec: conclusion}
We have presented a unified model-based statistical spatial downscaling framework that can be used in OSSEs to construct realizations of high-resolution NRs. Our model explicitly handles change-of-support that occurs because of the gap in spatial resolutions of the numerical model outputs and that of the desired NRs. Our downscaling framework differs from that in previous studies in two important ways. First, we only utilize numerical model outputs and do not require physical observations, which makes our method suitable in the context of OSSEs, in particular. Second, our downscaled NRs match the coarse-resolution outputs exactly when they are aggregated back to the resolution of the numerical model in order to preserve physical relationships embodied in the numerical model. 

We further proposed a data-driven algorithm to sequentially add basis functions to the low-rank component of the model to learn nonstationary spatial structures and localized features from data. This forward selection algorithm specifies both the centers and bandwidths of basis functions, adaptively.  When the number of observation locations is extremely large, we suggest that our algorithm be combined with spatial clustering methods such as \cite{Marchetti2017} to further improve computational efficiency. Our current algorithm selects the centers and bandwidths of basis functions in a forward fashion. One interesting extension would be a modification of such procedure to eliminate some basis centers by incorporating cross-validation steps in the algorithm.  LASSO-type variable selection methods \cite[e.g.][]{tibshirani1996regression, Bondell2010} can also be potentially developed for basis function selection, though more work is needed to make the classical regularization terms computationally practical for large spatial datasets.

In this article, we also show that by combining the low-rank and the CAR components together in the FGP model, the resulting downscaled field and spatial predictions are improved substantially. As in \cite{Ma2017}, alternative models that take into account the modeling and computational complexity trade-off can be adopted to describe spatial dependence in $\boldsymbol \xi$, such as the Gaussian Markov random field in \cite{Lindgren2011}. Expert knowledge of the likely behavior of the fine-scale process, if available, can also be incorporated into the model for $\boldsymbol\xi$. 

Our downscaling framework is designed to produce NRs at very fine spatial resolution. Ideally, we would like NRs not only at fine spatial resolution in a plane (horizontal resolution), but also at fine resolution in the vertical and temporal dimensions. As our downscaling framework is model-based, it can be extended to allow multiple input dimensions (longitude, latitude, height, time). The tensor basis functions in \cite{Nguyen2017} provide a way to describe both vertical and horizontal dependence. The spatio-temporal version of FGP model \citep{Ma2018DFGP} can be used to generate NRs across time. We will consider these extensions in future work.

The current downscaling framework can be used as a building block in hierarchical models for non-Gaussian distributions or nonlinear constraints. Compared to heuristic methods, our downscaling framework provides  a coherent and rigorous way to propagate physical relationships at coarse resolutions down to fine resolutions. Our current downscaling framework can also be extended to the multivariate downscaling framework to generate NRs for multiple geophysical processes. This is closely related to the theme of super-resolution imaging \citep{Tian2011}, which requires jointly downscaling multiple coarse-resolution images to obtain a single fine-resolution image.

Finally, our downscaling model is able to generate whole ensembles of high-resolution spatial fields through conditional simulation. These ensembles can facilitate probabilistic uncertainty quantification in observation system design and data assimilation algorithm evaluation at the fine resolutions at which those systems and algorithms are intended to operate. Meanwhile, our methods can be further extended to handle both numerical model output and physical observations, and thus may be useful in statistical emulation and uncertainty quantification for multi-fidelity computer models. It is also of interest to use the spatial modeling strategy in this article together with methods for computer model calibration to address a broad range of problems including data assimilation and retrievals in remote sensing and atmospheric sciences. These topics will be investigated in future research.

\section*{Acknowledgements}
The research was partially carried out at the Jet Propulsion Laboratory, California Institute of Technology, under a contract with the National Aeronautics and Space Administration. This material was based upon work partially supported by the National Science Foundation under Grant DMS-1638521 to the Statistical and Applied Mathematical Sciences Institute. Any opinions, findings, and conclusions or recommendations expressed in this material do not necessarily reflect the views of the National Science Foundation. This research was Ma's Ph.D. dissertation and was partially supported by the Charles Phelps Taft Dissertation Fellowship at the University of Cincinnati. Kang's research was partially supported by the Simons Foundation's Collaboration Award (\#317298) and the Taft Research Center at the University of Cincinnati. We wish to acknowledge Dr.~Noel Cressie, Dr.~Matthias Katzfuss, and Dr.~Vineet Yadav for valuable discussions and suggestions on this work. We thank the Editor, Associate Editor and two anonymous referees for constructive comments and suggestions.

\appendix 
\section*{Appendix}
\section{The EM Algorithm for Downscaling Models with FRK and FGP} \label{appendix: EM}
In this section, we will give complete derivation of EM algorithms for FRK and FGP under downscaling framework in detail. Recall that the complete data log-likelihood function is given in Eq.~\eqref{eqn: complete data log-likelihood}. The EM algorithm consists of two steps: E-step and M-step, and these two steps are run iteratively starting with initial values until the EM algorithm converges. Given parameter estimates $\btheta_{[\ell]}$ in the $\ell$-th iteration of the EM algorithm, the conditional distribution of $\boeta$ given $\widetilde \bY$ is multivariate normal with mean $\bmu_{\boeta| \widetilde \bY, \btheta_{[\ell]}}$ and covariance matrix $\bSigma_{\boeta| \widetilde \bY, \btheta_{[\ell]}}$, which are 
\begin{eqnarray}
\bmu_{\boeta| \widetilde \bY, \btheta_{[\ell]}} & = & \bK_{[\ell]} (\bA \bS)^T (\bA \bSigma_{[\ell]} \bA^T)^{-1} [\widetilde \bY-(\bA \bX) \bbeta_{[\ell]}], \\
\bSigma_{\boeta| \widetilde \bY, \btheta_{[\ell]}} &=& \bK_{[\ell]} - \bK_{[\ell]}(\bA \bS)^T (\bA \bSigma_{[\ell]} \bA^T)^{-1} (\bA \bS) \bK_{[\ell]}^T,
\end{eqnarray}
where the subscript ``$[\ell]$'' indicates that the quantity is evaluated with parameters $\bbeta_{[\ell]}, \bK_{[\ell]}, \sigma^2_{\xi, {[\ell]}}$ in FRK model, and with parameters $\bbeta_{[\ell]}, \bK_{[\ell]}, \tau^2_{[\ell]}, \gamma_{[\ell]}$ in FGP model. In E-step, taking conditional expectation of the complete data log-likelihood w.r.t.~$\boeta$ given $\widetilde \bY$ with parameters $\btheta_{[\ell]}$ will give the $Q(\btheta; \btheta_{[\ell]})$ function. The twice-negative $Q$ function is 
\begin{eqnarray} \label{eqn: Q function} \nonumber 
-2Q(\btheta; \btheta_{[\ell]}) &=& E_{\boeta|\widetilde \bY, \btheta_{[\ell]}}[-2\ln L(\boeta, \widetilde \bY)] \\ \nonumber 
&=& \ln | \bK| + \ln | \bD^{-1} |  + [\widetilde \bY-(\bA \bX) \bbeta]^T \bD [\widetilde \bY - (\bA \bX) \bbeta] \\ \nonumber
&-& 2 [\widetilde \bY-(\bA \bX)\bbeta]^T \bD (\bA \bS) \bmu_{\boeta| \widetilde \bY, \btheta_{[\ell]}} + tr\{ [(\bA \bS)^T \bD (\bA \bS) + \bK^{-1}] \bSigma_{\boeta| \widetilde \bY, \btheta_{[\ell]}} \} \\ \nonumber 
&+& \bmu_{\boeta| \widetilde \bY, \btheta_{[\ell]}}^T \bSigma_{\boeta| \widetilde \bY, \btheta_{[\ell]}} \bmu_{\boeta| \widetilde \bY, \btheta_{[\ell]}}.
\end{eqnarray}
In M-step, the $Q$ function is maximized w.r.t.~parameters $\btheta$ to obtain updated parameters $\btheta_{[\ell+1]}$. As the formulas for FRK and FGP are slightly different, we first give formulas for parameter updates in FRK. In the downscaling model based on FRK, taking derivative of $-2Q(\btheta; \btheta_{[\ell]})$ w.r.t. $\bbeta, \bK, \sigma^2_{\xi}$ and setting it to zero will give 
\begin{eqnarray} \label{eqn: update beta in FRK}
\bbeta_{[\ell+1]} &=& [(\bA \bX)^T (\bA \bA^T)^{-1} (\bA \bX)]^{-1} (\bA \bX)^T (\bA \bA^T)^{-1} [\widetilde \bY - (\bA \bS) \bmu_{\boeta| \widetilde \bY, \btheta_{[\ell]}}], \\
\label{eqn: update K in FRK}
\bK_{[\ell+1]} &=& \bSigma_{\boeta| \widetilde \bY, \btheta_{[\ell]}} + \bmu_{\boeta| \widetilde \bY, \btheta_{[\ell]}} \bmu_{\boeta| \widetilde \bY, \btheta_{[\ell]}}^T, \\ 
\label{eqn: update sigma_xi in FRK}
\sigma^2_{\xi, [\ell+1]} &=& \{ [\widetilde \bY-(\bA \bX) \bbeta_{[\ell+1]}]^T (\bA \bA^T)^{-1} [\widetilde \bY - (\bA \bX) \bbeta_{[\ell+1]} - 2 (\bA \bS) \bmu_{\boeta| \widetilde \bY, \btheta_{[\ell]}}] \\ \nonumber 
&+& tr[ (\bA \bS)^T (\bA \bA^T)^{-1} (\bA \bS) \bSigma_{\boeta| \widetilde \bY, \btheta_{[\ell]}}] \}/ M - \sigma^2_{\epsilon}. 
\end{eqnarray} 
In the downscaling model based on FGP, taking derivative of $-2Q(\btheta; \btheta_{[\ell]})$ w.r.t.~$\bbeta, \bK$, and setting it to zero will give  
\begin{eqnarray} \label{eqn: update beta in FGP}
\hat{\bbeta} &=& [(\bA \bX)^T \bD (\bA \bX)]^{-1} (\bA \bX)^T \bD [\widetilde \bY - (\bA \bS) \bmu_{\boeta| \widetilde \bY, \btheta_{[\ell]}}], \\
\label{eqn: update K in FGP}
\bK_{[\ell+1]} &=& \bSigma_{\boeta| \widetilde \bY, \btheta_{[\ell]}} + \bmu_{\boeta| \widetilde \bY, \btheta_{[\ell]}} \bmu_{\boeta| \widetilde \bY, \btheta_{[\ell]}}^T,
\end{eqnarray} 
where $\bK_{[\ell+1]}$ is updated explicitly, but $\hat{\bbeta}$ depends on values of $\tau^2$ and $\gamma$.  To get parameters updates $\tau^2_{[\ell+1]}$ and $\gamma_{[\ell+1]}$, the following function needs to be minimized wr.t.~$\tau^2, \gamma$: 
\begin{eqnarray} \label{eqn: update tau2, gamma}
f(\tau^2, \gamma) &=& \ln |\bD|  + [\widetilde \bY - (\bA\bX)\hat{\bbeta}]^T \bD [\widetilde \bY -(\bA\bX)\hat{\bbeta}]\\ \nonumber 
 &-& 2 [\widetilde \bY-(\bA\bX)\bbeta]^T \bD (\bA \bS) \bmu_{\boeta| \widetilde \bY, \btheta_{[\ell]}} + tr[ (\bA \bS)^T \bD (\bA \bS) \bSigma_{\boeta| \widetilde \bY, \btheta_{[\ell]}}],
\end{eqnarray}
where $\bD=[ \bA (\mathbf{I} - \gamma \bH)^{-1} \bA^T /\tau^2 + \bA \bA^T/\sigma^2_{\epsilon}]^{-1}$. By plugging in the function $f(\tau^2, \gamma)$ with $\hat{\bbeta}$ in Eq.~\eqref{eqn: update beta in FGP}, numerical optimization such as interior-point or active-set algorithm can be used to obtain optimal values for $\tau^2$ and $\gamma$. The optimal values $\tau^2_{[\ell+1]}$ and $\gamma_{[\ell+1]}$ are then plugged in Eq.~\eqref{eqn: update beta in FGP} to obtain parameter updates $\bbeta_{[\ell+1]}$. The function $f(\tau^2, \gamma)$ can be evaluated efficiently as it has same computational cost to evaluate the twice negative marginal log-likelihood function \eqref{eqn: likelihood fun}. To accelerate the EM algorithm in FGP, we use Akein's acceleration scheme to update EM algorithm, which is called SQUAREM algorithm in \cite{Berlinet2007} and \cite{Varadhan2008}.  

The initial value for $\bbeta$ in the EM algorithms of FRK and FGP can be set as the ordinary least square estimate $\hat{\bbeta}_{\text{ols}}=[(\bA\bX)^T(\bA\bX)]^{-1}(\bA\bX)^T\widetilde \bY$. The initial value for $\bK$ can be set to $0.9 \hat{\sigma}^2_{\widetilde\bY} \mathbf{I}_r$, where $\hat{\sigma}^2_{\widetilde\bY}$ is the empirical variance of $\widetilde \bY$. The initial value for $\sigma^2_{\xi}$ is $0.1\hat{\sigma}^2_{\widetilde\bY}$ in FRK. The initial value for $\tau^2$ can be set to $0.1\hat{\sigma}^2_{\widetilde\bY}$, and $\gamma$ is constrained in the interval $(1/\lambda_1, 1/\lambda_N)$, where $\lambda_1, \lambda_N$ are the samellest and largest eigenvalues for the proximity matrix $\bH$. The EM algorithm starts with the initial values $\btheta_{[\ell]}$ at $\ell=0$, and then the E-step and M-step are carried out iteratively with new initial values from previous M-step until certain convergence criterion is satisfied, e.g., the difference of the parameters $\btheta$ at two consecutive iterations is less than a threshold. The convergence of the EM algorithms is monitored by the twice-negative-marginal-log-likelihood function \eqref{eqn: likelihood fun}.

\section{Technical Proofs} \label{appendix: proof}

\noindent\textbf{\em Proof of Proposition~\ref{Property of conditional simulation}:}
\begin{itemize}[noitemsep,topsep=4pt]
\item[(1)] Recall the definition of $\bY_{\text{CS}}$ in Algorithm~\ref{algorithm:cs}, 
$\bY_{\text{CS}}= \bY_{\text{NS}} + \bSigma \bA^T (\bA \bSigma \bA^T)^{-1} (\bA{\bY} - \bA \bY_{\text{NS}}).$
It follows immediately that $\bA \bY_{\text{CS}} = \bA \bY_{\text{NS}} + \bA \bSigma \bA^T (\bA \bSigma \bA^T)^{-1} (\bA{\bY} - \bA \bY_{\text{NS}}) = \bA \bY.$ Thus, conditional on $\bA\bY=\widetilde\bY$, we have $\bA\bY_{\text{CS}}=\widetilde\bY$. 
\item[(2)] Let $\bH \equiv \bSigma \bA^T ( \bA \bSigma \bA^T)^{-1}$. Then $\bY_{\text{CS}}=\bY_{\text{NS}} + \bH(\bA \bY - \bA \bY_{\text{NS}})$. The expectation of $\bY_{\text{CS}}$ given $\bA\bY$ is 
\begin{eqnarray*}
E[\bY_{\text{CS}}\mid \bA\bY] &=& E(\bY_{\text{NS}}) + \bH [\bA \bY - \bA E(\bY_{\text{NS}})]\\
&=& \bmu + \bH (\bA \bY - \bA \bmu),\\
&=&\bmu+\bSigma\bA^T(\bA\bSigma\bA^T)^{-1}(\bA\bY-\bA\bmu).
\end{eqnarray*} 
The covariance matrix of $\bY_{\text{CS}}$ given $\bA\bY$ is 
\begin{eqnarray*}
\text{cov}(\bY_{\text{CS}} \mid \bA \bY) &=& \text{cov}( \bY_{\text{NS}} - \bH \bA \bY_{\text{NS}} \mid \bA \bY ) = \text{cov}[ (\mathbf{I} - \bH \bA) \bY_{\text{NS}} \mid \bA \bY ]\\
&=& (\mathbf{I} - \bH\bA) \text{cov}(\bY_{\text{NS}}) (\mathbf{I} - \bH\bA)^T 
= (\mathbf{I} - \bH\bA) \bSigma (\mathbf{I} - \bH\bA)^T \\
&=& \bSigma - \bSigma \bA^T \bH^T - \bH \bA \bSigma + \bH \bA \bSigma \bA^T \bH^T\\
&=& \bSigma - \bSigma \bA^T(\bA\bSigma\bA^T)^{-1} \bA \bSigma.
\end{eqnarray*}
Since $\bY_{\text{NS}}$ follows the multivariate normal distribution and $\bY_{\text{CS}}$ is a linear transformation of  $\bY_{\text{NS}}$ (conditional on $\bA\bY$), it is easy to verify that $\bY_{\text{CS}}\mid \bA\bY$ follows a multivariate normal distribution with mean and covariance given above. 
\item[(3)] It is obvious that $\bY_{\text{CS}}$ follows a multivariate normal distribution since it is a linear combination of multivariate normal vectors $\bY$ and $\bY_{\text{NS}}$. Therefore, it suffices to show that $\bY_{\text{CS}}$ has mean $\bmu$ and covariance matrix $\bSigma$. It follows that 
\begin{eqnarray*}
E(\bY_{\text{CS}}) = E[E(\bY_{\text{CS}}| \bA \bY)] 
= \bmu + \bH [\bA E(\bY) - \bA \bmu] = \bmu.
\end{eqnarray*}
The covariance matrix of $\bY_{\text{CS}}$ is 
\begin{eqnarray*}
\text{cov}(\bY_{\text{CS}}) &=& \text{cov}(\bY_{\text{NS}} - \bH \bA \bY_{\text{NS}} + \bH \bA \bY) 
= \text{cov}[(\mathbf{I} - \bH \bA) \bY_{\text{NS}}] + \text{cov}(\bH \bA \bY) \\
&=& (\mathbf{I} - \bH \bA) \bSigma (\mathbf{I} - \bH \bA)^T + \bH \bA \bSigma (\bH \bA)^T \\
&=& \bSigma - \bSigma \bA^T(\bA\bSigma\bA^T)^{-1} \bA \bSigma + \bSigma \bA^T(\bA\bSigma\bA^T)^{-1} \bA \bSigma\\
& = &\bSigma.
\end{eqnarray*}
\item[(4)] It follows that
$E(\bY_{\text{CS}} - \bY)^2 = \text{cov}(\bY_{\text{CS}} - \bY) 
= \text{cov}(\bY_{\text{CS}}) - \text{cov}(\bY_{\text{CS}}, \bY) -  \text{cov}(\bY, \bY_{\text{CS}}) + \text{cov}(\bY)$, which is $2[\bSigma - \bSigma \bA^T (\bA\bSigma \bA^T)^{-1} \bA \bSigma].$

\end{itemize}

\bibliographystyle{apa}

\singlespacing 
\setlength{\bibsep}{5pt}

\bibliography{downscaling_ref}

\end{document}